\documentclass[12pt,preprint]{aastex}

\def\kms{km ${\rm s}^{-1}$}
\def\sn1{${\rm s}^{-1}$}
\def\cmd{${\rm cm}^{-3}$}
\def\cmn{${\rm cm}^{-2}$}
\def\ergs{erg ${\rm s}^{-1}$}
\def\flux{ergs cm$^{-2}$ ${\rm s}^{-1}$}
\def\mr{MR~2251$-$178}

\def\sp{\space}

\shorttitle{{\it Chandra} LETGS spectroscopy of ionized absorbers}
\shortauthors{Ram{\'i}rez et al.}

\begin{document}

\bibliographystyle{apjsty}

\title{{\it Chandra} LETGS spectroscopy of the Quasar \mr \space and its warm absorber}

\author{
J.M.~Ram{\'i}rez,   
Stefanie~Komossa,   
Vadim~Burwitz   
}

\affil{Max-Planck-Institut f\"{u}r extraterrestriche Physik, D-85741 Garching, Germany}
  
\and
\author{ Smita~Mathur}   

\affil{Department of Astronomy, Ohio State University, 140 West 18th Avenue,
  Columbus, OH 43204}


\begin{abstract}

We present an analysis of our {\it Chandra}
Low Energy Transmission
Grating Spectrometer
(LETGS) observation of the quasar \mr.
The warm absorber of \mr \space is well described by a
hydrogen column density, $N_{\rm H}\approx 2\times 10^{21}$ \cmn,
and an ionization parameter $\log(\xi)\approx 0.6$.
We find in the spectrum weak evidence for
narrow absorption lines from Carbon and Nitrogen
which indicate that the ionized
material is in outflow.
We note changes (in time) of the absorption structure
in the band ($0.6-1$) keV
(around the UTAs plus
the O~{\sc vii}
and O~{\sc viii} K-edges) at different periods of the observation.
We measure
a ($0.1-2$) keV flux
of 2.58 $\times 10^{-11}$ \flux.
This flux
implies that the nuclear source of \mr \space
is in
a relatively low
state. No significant variability
is seen in the light curve. We do not find evidence
for an extra cold material in the line of sight,
and set an upper limit of
$N_{\rm H} \approx 1.2 \times 10^{20}$ \cmn.
The X-ray spectrum does not appear to show evidence
for dusty material, though an upper limit in the neutral carbon
and oxygen column densities can only
be set to
$N_{\mbox{{\tiny C~I}}}\approx 2 \times 10^{19}$ \cmn \space and
$N_{\mbox{{\tiny O~I}}}\approx 9 \times 10^{19}$ \cmn, respectively.

\end{abstract}

\keywords{galaxies: active --
X-rays: galaxies -- quasars: individual (\mr)}


\section{Introduction}

Warm absorbers
have provided deep insights into the nuclear
environment of Active Galaxies (AGNs). 
\cite{halpern1984a} reported its presence
for the first time using the {\it Einstein} observation of the 
QSO \mr. Since then, warm absorbers have been commonly found in about
50 \% of the AGN spectra
\citep[see][ for reviews]{komossa2003a,crenshaw2003a},
and their study has enriched our
knowledge about the ionization and the kinematics of the
gas composing these systems, important for the understanding
of the evolution of these objects, and the AGN 
unification picture.

In this context \mr,
at $z=0.06398$ \citep{bergeron1983a},
has been the subject of various studies,
first due to its historical importance as the first 
quasar discovered by X-ray observations 
\citep{ricker1978a},
the first warm absorber
reported \citep{halpern1984a}, 
and because it displays a number of 
outstanding characteristics.
It is surrounded by a large
[O~{\sc iii}] emission line region,
located in the outskirts of a cluster
of galaxies \citep{bergeron1983a},
with a high ratio 
L$_{\rm x}$/L$_{\rm opt}$.
Several studies have been focused on the
characterization of the
absorbing material properties.
\cite{komossa2001a} fitted a warm absorber
model to the {\it ROSAT} observation
of \mr, and found a high ionization
state of the absorber 
($\log U \sim 0.5$),
and a column density
of $N_{\rm H}\sim 10^{22.6}$ \cmn.
\cite{monier2001a} reported the presence of 
absorption lines of N~{\sc v}~$\lambda 1240$ and
C~{\sc iv}~$\lambda 1549$ blueshifted by few hundreds of
\kms. \cite{ganguly2001a} noticed variability
of the C~{\sc iv}~$\lambda 1549$ line by taking observations
from the {\it Hubble Space Telescope} (HST)
4 years apart, and inferred a maximum
distance from the source of $\sim$ 2.4 kpc.

More recently, \cite{kaspi2004a} established a scenario
where clouds
crossing
our line
of sight show up by the presence of 
absorption lines coming
from high
ionization
states, displaying a wide velocity range
from 0 to $\sim 600$ \kms, taking a 8.5 yr set of data
from {\it ASCA}, {\it BeppoSAX} and {\it XMM-Newton}.

\cite{gibson2005a} analyzed data from the
{\it High Energy Transmission Grating} on board $Chandra$,
and found evidence in the spectrum of \mr \space
of highly ionized, high-velocity
($v\sim 12,000-17,000$ \kms) outflowing material.
They report similar column densities to those
of \cite{kaspi2004a}, of a few 10$^{21}$ \cmn,
and
establish conditions for the accretion 
and mass-loss rates.

From the Fe~{\sc xxvi} L$_{\alpha}$ line,
\cite{gibson2005a} concluded that, unless
the absorber covering factor is very low,
the mass-loss rate is approximately
an
order of magnitude
higher than the accretion rate that would provide
the radiation power of \mr
\space ($\sim 0.2$ $M_{\odot}~{\rm yr}^{-1}$,
at 10 \% efficiency). This is a different
conclusion from that reached by \cite{monier2001a},
in the {\it Hubble Space Telescope}
(HST) analysis of the same object, where
they established that the accretion and mass-loss
rates are essentially the same. 
The inquiry into this point
is of relevance to the
understanding of the evolution mechanisms in
the nucleus of \mr.
On the other hand, the HETGS
high resolution observation
also reveals the presence of emission lines
coming from highly-ionized species, and
shows that the emitting material is not in our
line of sight, opening the possibility that 
the
absorption and the emission lines have different
physical origins or that
they are located at different
distances from the nucleus. This phenomenon has been
observed in NGC 4051 in the optical/ultraviolet band.
\cite{komossa1997a} concluded that the coronal
emission lines from lowly ionized species
of Fe ([Fe~{\sc vii}]-[Fe~{\sc xi}]) observed in 
the spectrum of this
object, have a {\it different physical origin} from the warm absorber,
consistent with the fact that they could come from
a different spatial region,
the warm absorber located at the nucleus and the
coronal lines extended out of
 $\sim 150$~pc
\citep{nagao2000a}.

Finally dusty warm absorbers
have been found in a number of AGN 
\citep[][ and references therein for a review]{komossa2003a}
and are predicted to leave
their mark on the X-ray spectra of AGN,
for instances the K-edges of O~{\sc i} and C~{\sc i}
\citep{komossa1997a,komossa1998a}, and the Fe-L edge
\citep{lee2001a}.
For the first time, we search for the presence of a dusty
warm absorber in \mr. Based on ROSAT data, \cite{komossa2001a} speculated
about the presence of a second dusty warm absorber in \mr, this second warm absorber dusty,
but the data
did not allow multi-component fitting,
which would provide
important
implications on the global structure of \mr.

It is clear from this introduction that
each of the previous missions and the use of
more powerful spectroscopic instruments
lead to new perspectives regarding the
properties and the evolution of the
nuclear environment of \mr.
We present an overview and
an analysis of the spectrum of \mr \space
as it
is seen with the {\it Low Energy Transmission
Grating Spectrometer} (LETGS) on board {\it Chandra}.
The spectral resolution
power of the instrument allows us
to confirm some of the previous conclusions
about the ionization and the kinematics of the warm
absorber of \mr, and
shed new light on the evolution, ionization
and composition
of this system.

Technical details 
on the observation and data reduction
are given in \S \ref{obs11}.
Our spectral analysis is given
in \S \ref{models11}.
The mean properties of the warm absorber of \mr \space
are presented in \S \ref{xstar_calc1},
with the corresponding spectral absorption
lines 
analysis.
To finish, we made a temporal analysis
on the absorption structure around the
O~{\sc vii} and O~{\sc viii} K-edges
in \S \ref{temp11}. We discuss the results in \S \ref{discuss11}
and conclude in \S \ref{conclu11}.
Throughout this paper, we use the following cosmological parameters:
H$_0=70$ \kms \space Mpc$^{-1}$,
$\Omega_{\rm M}=0.3$ and $\Omega_{\lambda}=0.7$.


\section{Observation and Data Reduction
\label{obs11}}

We obtained a $\sim$ 80 ks exposure time
observation of \mr, performed with the
{\it Low Energy Transmission
Grating Spectrometer} \citep[LETGS,][]{brinkman2000a}
on board {\it Chandra}
(under the sequence number 700405 and OBS ID 02966).
A log of the observation is presented in Table
\ref{tbl1}.
The spectrum was obtained by reducing the data
with the Chandra Interactive Analysis of Observation
(CIAO\footnote{\tt http://cxc.harvard.edu/ciao})
version 3.3.

We use the default spectral extraction region
(i.e., a
``bow-tie" shaped region).
When the LETG 
({\it Low Energy Transmission
Grating}) is used with the HRC-S detector 
({\it High Resolution Camera}), this
comprises a central rectangle
abutted to outer regions whose widths increase
as the dispersion distance increases.
The background region is taken from
above and below the dispersed spectra.
The region shape for both, source and background
negative and positive orders is precisely given
in the file
{\tt letgD1999-07-22regN0002.fits},
that can be found 
at the LETGS calibration webpage
\footnote{{\tt http://cxc.harvard.edu/cal/Letg}}.

Having properly extracted both, source and background
spectra from each arm, we merged them, obtaining added
source and background spectra. This is intended
to increase the 
signal-to-noise ratio
(S/N) of the final spectrum, and the spectral
analysis (throughout this work) is based on this
co-added spectrum.
The effective areas (EA) for orders from 2 to 10
used in fitting procedures were
taken from the LETG+HRC-S effective areas webpage
\footnotemark[2].
For the first order we used the corrected
EA of \cite{beuermann2006a}.
EA orders from 1 to 10 (positive and negative orders) 
were summed to be used with the corresponding co-added 
spectrum.

\begin{figure}
\rotatebox{0}{\resizebox{15cm}{!}
{\plotone{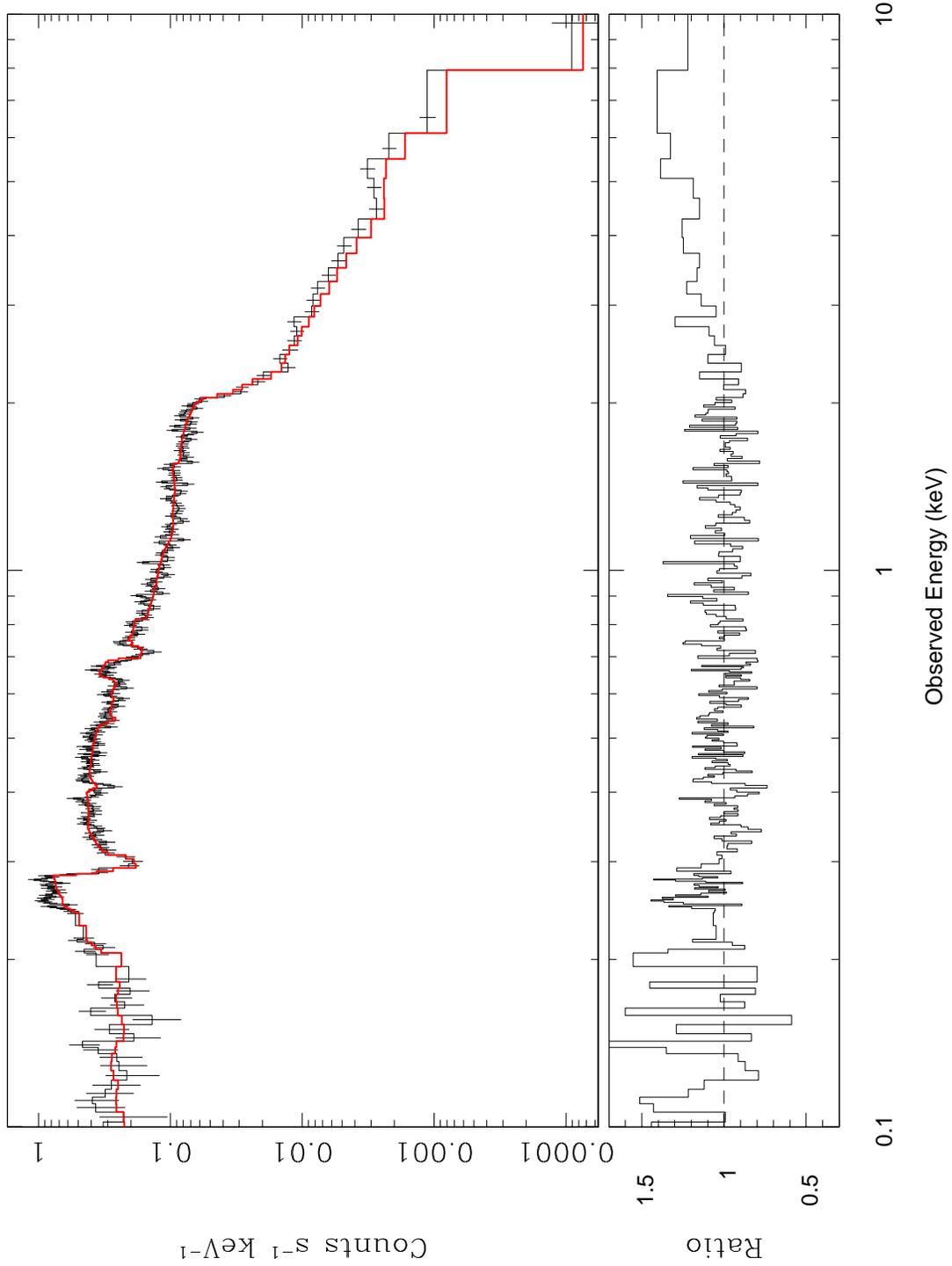}}}
\caption{
0.1-10 keV LETGS spectrum of \mr
\space (observed frame).
It is
binned to have at least 100 counts from the source per bin.
In red the model of a simple absorbed power-law modified by 2 oxygen absorption
edges.
\label{full_range1}}
\end{figure}

The nominal LETGS wavelength range is $1.24-175$ ${\rm \AA}$ ($0.07-10$ keV).
However, we restrict our spectral analysis to the range $1.24-124$ ${\rm \AA}$
($\sim$ $0.1-10$ keV). In the range $\sim$ 0.07$-$0.09 keV the
S/N ratio is low ($\approx$ 1.4 at $\sim 0.09$ keV), and we exclude bins in this
band from our analysis. Also, we exclude bins that fall in the bands where the
gaps of the detector (HRC-S) are located (i.e., 52$-$56 \AA \sp and 62$-$66 \AA, for left
and right gaps respectively).
Figure \ref{full_range1} shows the summed
(positive and negative dispersion orders)
background subtracted count rate spectrum of \mr \space adaptively binned to have
at least 100 counts from the source
per bin over the range considered in this work.
It is known that an instrumental feature $\sim 2.08$ keV, related to
the mirror, could be modeled by an edge with negative optical depth
\footnote{{\tt http://cxc.harvard.edu/ccw/proceedings/03\_proc/presentations/marshall2/s001.html}}.
We model this Ir M edge of the mirror by using an edge model with energy
fixed to 2.08 keV, and an optical depth equal to $-0.15$.
So all our fits include this negative edge model.
All wavelengths and energies are presented in the observed frame.
\subsection{Source Extent and Light Curve
\label{lc1}}

In order to search for evidence of a spatial extent
of the source or the presence of
other possible X-ray sources, for example jets, close
to \mr,
we take
the zero-order image
(Chip 2 on the HRC-S) of \mr \space
centered at
RA=~22:54:05.08 and
Dec(J2000)=~-17$^{\circ}$34$^{'}$55$^{''}$
in a field of view of
10$^{''}\times$ 10$^{''}$.
The extent of the emission
(and the 90 \% encircled energy;
computed with the {\tt celldetect} command of CIAO\footnotemark
\footnotetext{{\tt http://cxc.harvard.edu/ciao3.3/download/doc/detect\_manual/cell\_run.html}})
is consistent with being inside the
point-spread function (PSF) dominated zone
(for the LETGS $\sim 1.8 ^{''}$).
We conclude that no evidence of appreciable
extent of the source is found in this observation
(we also checked for angular dependence in the X-ray image, and found
that the emission of the object is symmetrically distributed).

\begin{figure}
\rotatebox{0}{\resizebox{15cm}{!}
{\plotone{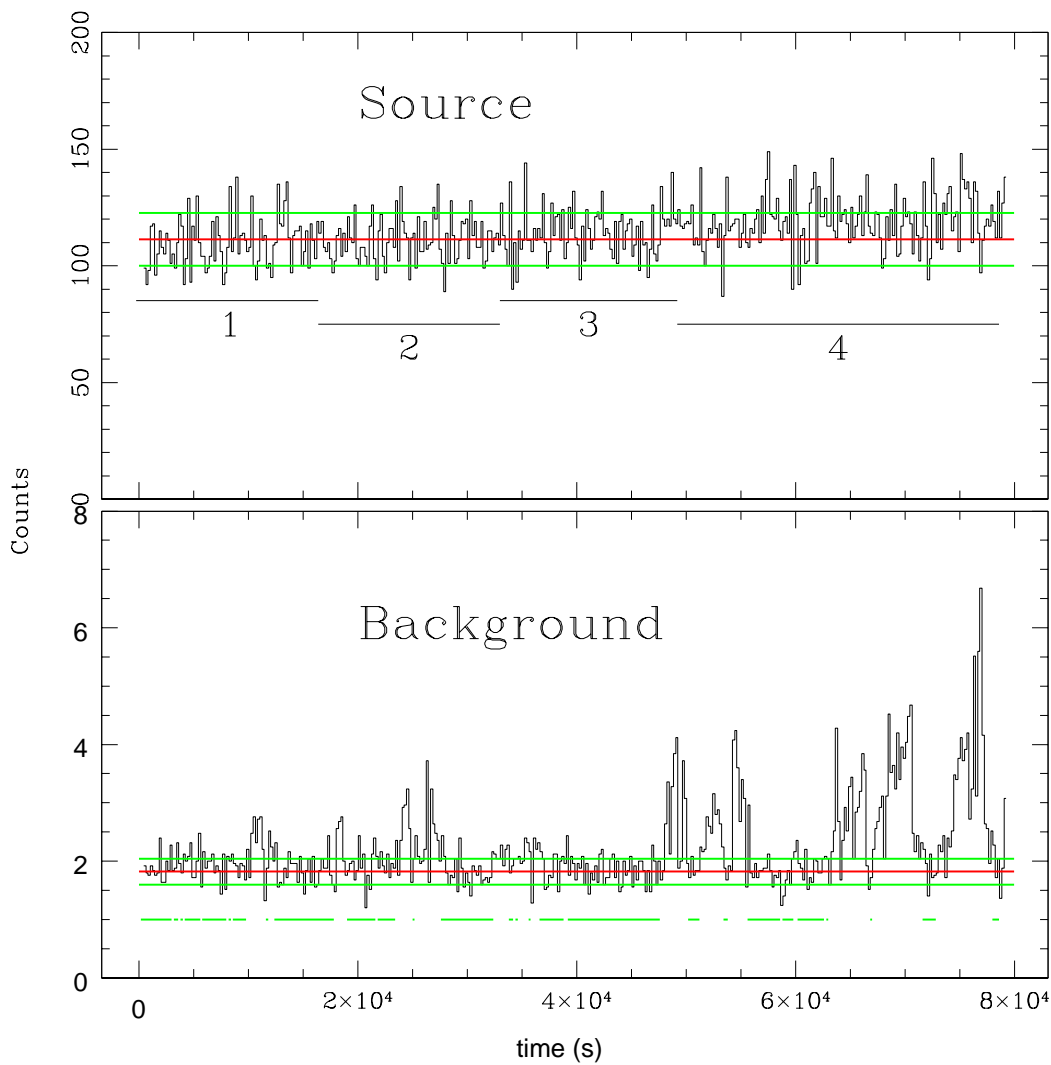}}}
\caption{
Light curves of the source and background (bin size 200 s).
The top panel shows, in red, the mean counts of
the source. No significant time variability is seen within
$\pm 1 \sigma$ (green lines). The  bottom panel shows
strong variability of the background
(here, the green lines show $\pm 3 \sigma$).
The good time intervals
are indicated by the interrupted line below the background
spectrum.
The meaning of intervals (1)-(4) is explained in
\S \ref{temp11}.
\label{light1}}
\end{figure}

The light curve constructed from the zero-order of the
observation is shown in Figure \ref{light1}
(top panel). 
Counts are binned to 200 s bins.
We find a mean value of 111 counts bin$^{-1}$
(horizontal red line) without any significant variation
within 1$\sigma$ (horizontal green lines). In the
bottom panel we present the light curve of the background
(see definition below), which shows that 
strong count variations occur
in several time intervals. Data from these intervals were
excluded from the spectral analysis presented below.
We removed data coming from time intervals whose count
rate were above 3$\sigma$ (horizontal green lines)
from the mean value (horizontal red line). The final spectrum
was constructed from these good time intervals (GTI),
time intervals where counts are within 3$\sigma$ from the mean
value of the fluctuation exhibited by the background.
A circular region around the zero-order image
with a radius of 10$^{''}$ was used to extract
the source light curve.
The background extraction region is defined 
by four
circles with radii of 25$^{''}$ equidistantly distributed
from the source in order to avoid contamination from events
associated with dispersion and cross-dispersion directions.

\section{Global Spectral Models
\label{global11}}

\subsection{Flux and Luminosity ($0.1-10$ keV)
\label{flux11}}

To compute the hard, and soft X-ray fluxes we
fit a simple power-law to the
total
spectrum\footnote{The added spectrum described in
section \ref{obs11} adaptively binned to have at least
100 counts from the source per bin.
The fit was carried
out in XSPEC 11, and the fluxes and luminosities computed
with the commands {\tt flux} and {\tt lumin} respectively
(the used band was $0.1-10$ keV, and the model includes the
Ir edge at 2.08 keV with $\tau=-0.15$).
We note that the LETGS spectrum of \mr \sp is 
dominated by the soft X-rays band (i.e, $< 2$ keV). So there might be an underestimate of the flux in the band
$2-10$ keV. The error can be estimated around 20 \%.}
absorbed only by the 
Galactic 
column density of $N_{\rm H}=
2.77 \times 10^{20}$ \cmn~\citep{lockman1995a},
and we obtain a photon index of
$\Gamma_{\rm x}=1.94 _{-0.03} ^{+0.03}$ and a normalization
value of 5.64 $_{-0.08} ^{+0.08}\times 10^{-3}$ photons
keV$^{-1}$ cm$^{-2}$ s$^{-1}$ at 1 keV.
Then we integrate the unabsorbed power-law
\footnote{We also compute the flux using a two powerlaws model, which represents in a better way the $2-10$ keV band,
with deviations from the data not larger than 10\%. The resulted fluxes do not change significatively.}
to obtain a flux
in the $0.1-2$ keV band
$f_{(0.1-2~{\rm keV})}=2.58 _{-0.03} ^{+0.04}\times 10^{-11}$ \flux,
and the ($0.1-2$) keV luminosity is
$L_{(0.1-2~{\rm keV})}=2.54\times 10^{44}$ \ergs,
a factor of $\approx$ 1.5 fainter than the ROSAT soft X-ray observation.
The ($2-10$) keV flux is
$f_{(2-10~{\rm keV})}=1.64$ $_{-0.05} ^{+0.05}\times 10^{-11}$ \flux,
and a ($2-10$) keV source luminosity
$L_{(2-10~{\rm keV})}=1.62\times 10^{44}$ \ergs.
This is
$\approx 30$ \% lower than the $2-10$ keV luminosity
measured by \cite{gibson2005a} of
$L_{(2-10~{\rm keV})}=2.41\times 10^{44}$ \ergs \sp (but see footnote 5).
In our computation,
284 of the total of 321 bins included in this
analysis are in the $0.1-2$ keV band, so all the spectral analysis is actually
driven by the {\it soft} part of the spectrum.
The information contained in the $0.1-0.5$ keV band is
important to conclude that the soft X-ray
flux is between
$\sim$ $10-80$ \% of the hard fluxes historically
measured from the spectrum of \mr.
Recently, this object has been observed in the hard X-ray band (14$-$195 keV) 
with the $Swift$ Burst Alert Telescope (BAT), with a
$L_{(14-195~{\rm keV})}=10^{45}$ \ergs \sp \citep{markwardt2005a}.
Taking our best-fit power-law and integrating over the 14$-$195 keV band,
we predict a luminosity $L_{(14-195~{\rm keV})}=3\times 10^{44}$ \ergs,
a few factor lower than the $Swift$ observation.
The full range ($0.1-10$ keV) luminosity,
is
$L_{(0.1-10~{\rm keV})}=4.17\times 10^{44}$ \ergs.
Our measured ($0.1-2$) keV flux
allows us to conclude that \mr \space is in
a relatively low soft X-ray state.


\subsection{Spectral Models
\label{models11}}

The rest of the
spectral analysis of the present work was carried out with the
Interactive Spectral Interpretation System package
\footnote{{\tt http://space.mit.edu/CXC/ISIS/index.html}}
(ISIS ver. 1.4.4) using the XSPEC
modules from the {\sc lheasoft}
libraries\footnote{{\tt http://heasarc.nasa.gov/docs/software/lheasoft/}}.


\mr \sp is surrounded by a giant emission-line
nebula of $\sim 10-200$ kpc extent \cite[e.g.,][]{bergeron1983a,macchetto1990a}.
\cite{bergeron1983a} estimated a column density of a few
times $10^{21}$ \cmn \sp of the emission-line gas.
They speculated about the presence of an extended neutral H~{\sc i} halo
around \mr \sp which should be detectable in H~{\sc i} 21cm
observations, or in the soft X-ray band, if also located
along our line-of-sight \citep[see also Sect 6.1 of][ for further motivation]{komossa2001a}.
We begin our fitting procedure by searching for, and placing
constraints on, any  cold material along our line-of-sight
to the quasar which is in excess to the Galactic absorption
toward \mr.

\begin{figure}
\rotatebox{0}{\resizebox{15cm}{!}
{\plotone{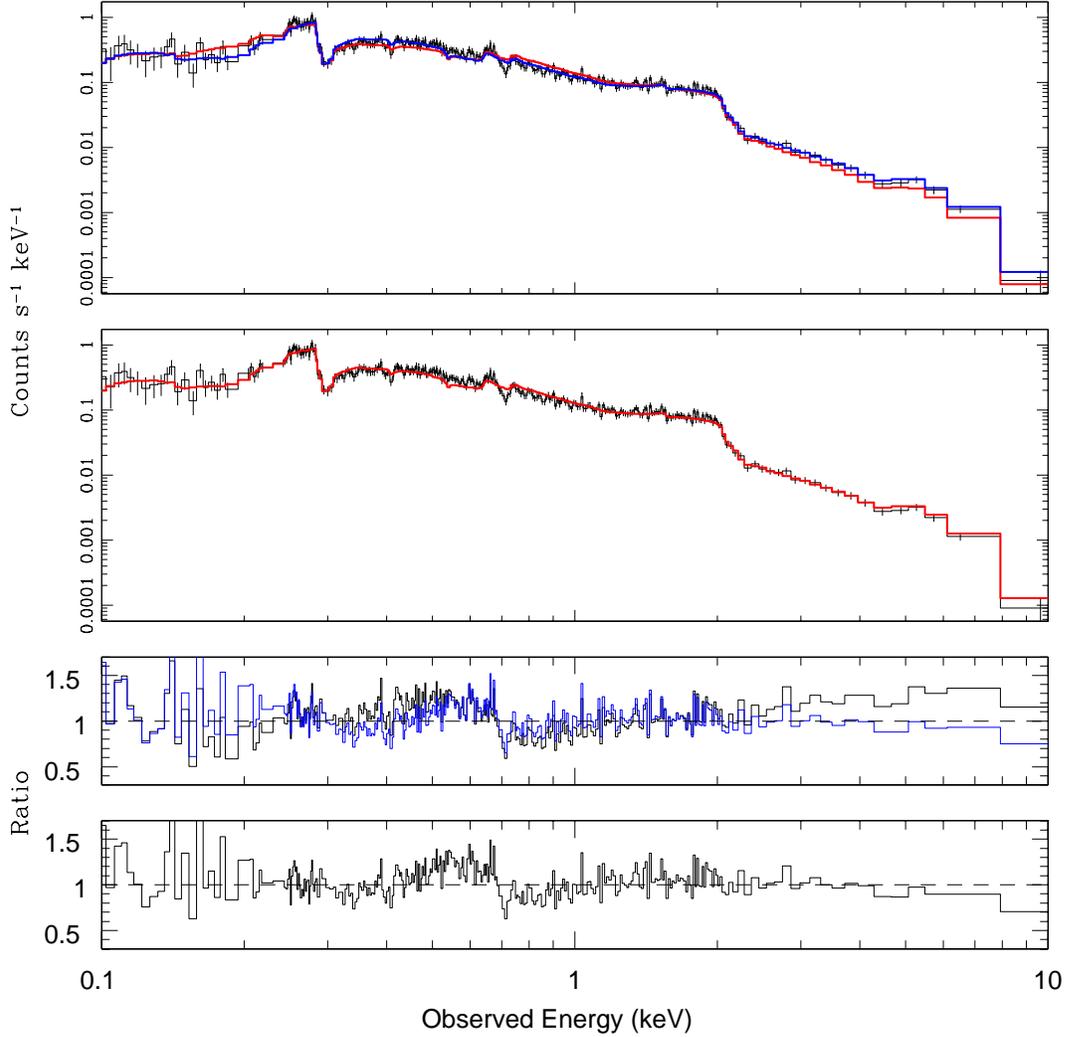}}}
\caption{
LETGS spectrum of \mr~(positive and negative orders added)
\space in energy space.
{\bf Top panel}: The solid red line shows the absorbed power-law fit.
The solid blue line is the absorbed {\tt pl+bb} model.
{\bf Middle panel}:
The solid line is the absorbed two power-law model.
{\bf Bottom panels}: Ratios of the composite models to the data:
absorbed power-law (black) and blackbody plus power-law (blue),
in the bottom panel the ratio for the two power-laws model
($absorbed$, means cold absorption with {\tt zphabs} ).
\label{cold1}}
\end{figure}

We fit a simple power-law, absorbed by a column of cold gas in the line of sight,
to the X-ray spectrum of \mr. Both, the power-law parameters and the column density
are free parameters. The best-fit
photon index and
column density we obtain are
$\Gamma_{\rm x}=1.79 _{-0.02} ^{+0.02}$ and 
$N_{\rm H}=1.68 _{-0.07} ^{+0.07} \times 10^{20}$ \cmn,
respectively. We plot in Figure
\ref{cold1} the best-fit absorbed power-law model along
with the spectrum of \mr \space in energy space. 
The best-fit column density is about
40 \% lower than the Galactic value
toward \mr \space \citep[2.77 $\times 10^{20}$ \cmn,][]{lockman1995a}.
If we fix the column density to the Galactic value and re-fit, we obtain
a slight change in the value of the photon index to
$\Gamma_{\rm x}\sim$ 1.9 and the
fit gets worse with a change in $\Delta \chi^2 \sim 140$.
Taking the upper
limit of the extra cold material reported by \cite{komossa2001a}
of $\Delta N_{\rm H}\equiv 5 \times 10^{19}$ \cmn \space and
including it in the column of gas,  we obtain a
photon index slightly
steeper and the fit gets worse if we increase the column density
of the extra cold material
by one and two times $\Delta N_{\rm H}$ (see rows 3 and 4 in Table \ref{tbl2}),
increasing $\chi^2$ (respect with $N_{\rm H}$ free to vary)
by $\Delta \chi^2 = 260$ and 396 respectively.

We also carefully investigate if excess X-ray absorption,
above the Galactic value,
and a blackbody
component could be combined to compensate each other
so as to mimic a less absorbed spectrum.
This case is shown graphically in Figure \ref{cold1} in solid
for the composite model
(i.e., {\tt zphabs $\times$ [pl+bb]}
\footnote{power-law plus blackbody modified by a column of gas in the AGN-frame.
The model is corrected by the Galactic absorption.})
in comparison with the
absorbed power-law (top panel).
The best-fit intrinsic (AGN frame) extra column density has
a value (given the data) consistent with zero.

We tried a last model
composed of two power-laws and cold absorption. The result is
shown in Figure \ref{cold1} (second panel
from top to bottom).
A steep power-law ($\Gamma_{\rm x}\approx 3.5$),
with help of the
cold absorption,
describes relatively well the soft band of the
spectrum ($\sim 0.1-0.6$ keV). A flatter second
power-law with $\Gamma_{\rm x}\sim 1$ is responsible for
describing the hard band ($\sim 0.7-10$ keV).
An extra absorption, above
the Galactic value, arises naturally from the fit
(with the same number of parameters
as the {\tt pl+bb} model). A material with column density
$N_{\rm H}\approx 1.2 \times 10^{20}$ \cmn \space
accounts for this extra absorption
(see the best-fit column density in Table \ref{tbl2} row 6).
We enforced this model to have an excess absorption of
this best-fit column density plus 1 and 2 times
$\Delta N_{\rm H}$, and re-fit, obtaining worse fits
(see rows 7 and 8 in Table \ref{tbl2}).
We conclude that there is little evidence
for cold excess absorption
with a column density of $N_{\rm H} \sim 10^{21}$ \cmn.
The maximum amount of cold absorption we can hide
in the X-ray spectrum of \mr, based on
the model fit involving two powerlaws,
is $N_{\rm H} \approx 1.2 \times 10^{20}$ \cmn \space,
and we consider this as a safe upper limit to the excess absorption
along the line of sight at the epoch of observation.

\begin{figure}
\rotatebox{0}{\resizebox{15cm}{!}
{\plotone{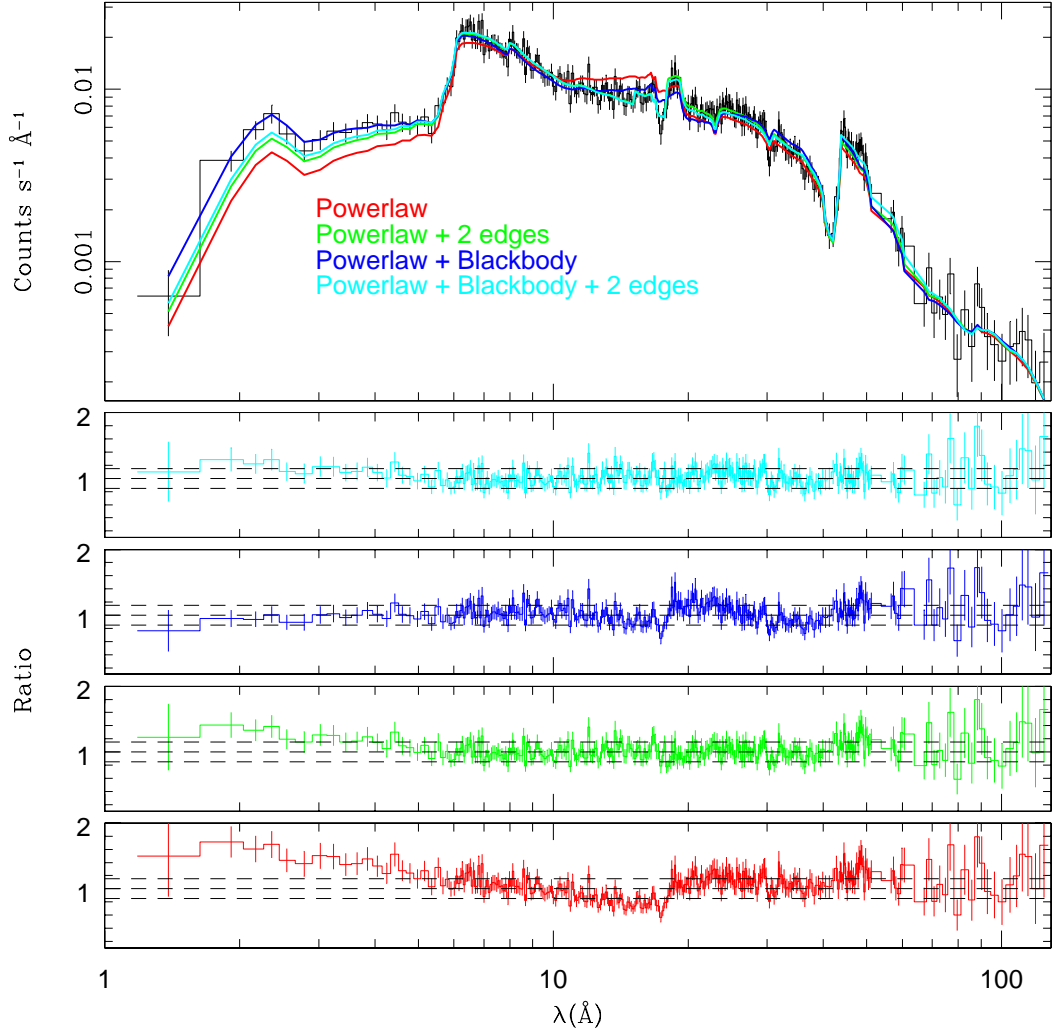}}}
\caption{
LETGS spectrum of \mr \space in the observed frame
(positive and negative orders added).
It is
binned to have at least 100 counts from the source per bin.
All the models include the Galactic absorption toward
\mr \space (see text). {\bf Top panel}: Solid lines:
{\tt pl} (red), {\tt pl $\times$ 2 edges} (green),
{\tt pl+bb} (dark blue),
{\tt (pl+bb) $\times$ 2 edges} (light blue).
The rest of the panels are the ratios of each composite model
to the data, represented by their corresponding colors
({\tt pl=power-law}; {\tt bb=blackbody}).
\label{conti_models1}}
\end{figure}

From now on, all the models considered in this study are
multiplied by an absorption column of gas
({\tt phabs} model in
XSPEC\footnote{{\tt http://heasarc.nasa.gov/docs/xanadu/xspec/manual}-
The abundances are solar from \cite{anders1989a}})
with $N_{\rm H}=2.77 \times 10^{20}$ \cmn  \space
to describe
the Galactic absorption toward \mr \space \citep{lockman1995a}.
First, we focus our attention on the fit of a power-law
to describe the 1.24$-$124 \AA \sp
band of the spectrum.
A single power-law gives
$\Gamma_{\rm x}=1.94 _{-0.02} ^{+0.02}$,
with large residuals
(i.e., 15$-$40 \%) in the $\sim$ 1.2$-$7 \AA \sp
and close to 18 \AA \sp
(see Figure \ref{conti_models1}),
and a poor statistical fit quality of
$\chi^2_{\nu}=3.00$.
We included a thermal component to account
for any soft X-ray excess. A blackbody
with a temperature of
$kT=81.5 _{- 2.0} ^{+ 2.0}$ eV
significantly improved the fit with
$\chi^2_{\nu}=1.63$
(for ${\rm d.o.f}=305$).
With this addition, the slope of the power-law changes
to a flatter value of $\Gamma_{\rm x}=1.53 _{-0.02} ^{+0.02}$, and
the model agrees with the data within 30 \%
almost from $\sim$ 1.2 to 40 \AA, except for
the residual around 18 \AA.

Now, let us consider the absorption features at $\approx$ 18 ${\rm \AA}$ and
$\approx$ 15 ${\rm \AA}$ in turn.
\cite{komossa2001a}  fitted a power-law with two edges at the theoretical
positions of O~{\sc vii} and O~{\sc viii}, 739~eV and 871~eV respectively,
representing the warm absorber, obtaining
$\tau_{\mbox{{\tiny O~VII}}}=0.22\pm 0.11$
and $\tau_{\mbox{{\tiny O~VIII}}}=0.24\pm 0.12$.
The high sensitivity and resolution
of the {\it Chandra} LETG spectrometer coupled with the good quality of
the observation of \mr, allow us to fit these features with a high
precision. Our {\tt pl$\times$ 2 edges} ($\approx$ 18 ${\rm \AA}$ and
$\approx$ 15 ${\rm \AA}$ in the observed frame)
model gives us a significant improvement
($\chi^2_{\nu}=1.38$
for 305 d.o.f)
over the fit of the power-law alone
($\chi^2_{\nu}=3.00$
for 307 d.o.f)
and the power-law plus the
blackbody component as well
($\chi^2_{\nu}=1.63$
for 305 d.o.f).
This model is in agreement with the
data $\approx \pm$ 15 \%
(data/model $\approx 1.00 \pm 0.15$,
Figure \ref{conti_models1})
over almost the entire spectrum, except for the
wavelength range ~1.2$-$6 ${\rm \AA}$
(and possibly at wavelengths $\gtrsim 60$ \AA \sp, but see
the large errors).
The continuum optical depths are
$\tau_{\mbox{{\tiny O~VII}}}=0.56 _{-0.04} ^{+0.05}$ and
$\tau_{\mbox{{\tiny O~VIII}}}=0.22 _{-0.04} ^{+0.05}$.
We compare these values with those
obtained from {\it ROSAT}.
In the case of
O~{\sc viii} they appear to agree within the error bars.
However, in the case of O~{\sc vii} the optical depth is
approximately a factor of 2.5 higher.
In this direct comparison model-to-model, the {\it Chandra} data
suggests that the contribution of this feature has
changed over the years ($\sim$ fifteen).

Finally, we can build a model combining all the ingredients mentioned
above leading us to a global view of the spectrum:
A power-law plus a soft-thermal component modified by two
warm absorber edges. This model is shown in
Figure \ref{conti_models1} and the maximum deviation of the
ratio data/model is $\pm$ 15 \% over
almost the entire spectrum, with $\chi^2_{\nu}=1.25$
(for 303 d.o.f). The F-test identifies this
improvement (the inclusion of the black body emission) as significant (at $>99$\% significance level),
over the {\tt pl$\times$ 2 edges} model.
The blackbody has a temperature
of $kT\sim$ 50 eV with the slope of the power-law being
in agreement with other models of this
study ($\sim 1.7-2$) and with slopes reported
by previous missions (i.e. {\it ROSAT}), but not with the flatter
slope of $\sim$ 1.4 reported recently by \cite{gibson2005a} in the HETGS
observation of the same object.
We can see that the inclusion of the strong O~{\sc vii} feature
displaces the temperature of the blackbody from
$kT\sim 80$ to 50 eV with
respect to the {\tt pl+bb} without edges.
The reduced temperature of the {\tt bb} helps to improve the fit
in the range $\sim 20$ to 60 \AA, giving an improvement
of $\Delta \chi^2 \approx 110$.
The model parameters are quoted in Table \ref{tbl3}.
The four models are plotted in Figure \ref{conti_models1} upon the
background subtracted
count rate spectrum of \mr \sp in the wavelength space 
(re-binned to have at least 100 counts from the source per bin).
\subsection{A Dusty Warm Absorber?}
\begin{figure}
\rotatebox{0}{\resizebox{15cm}{!}
{\plotone{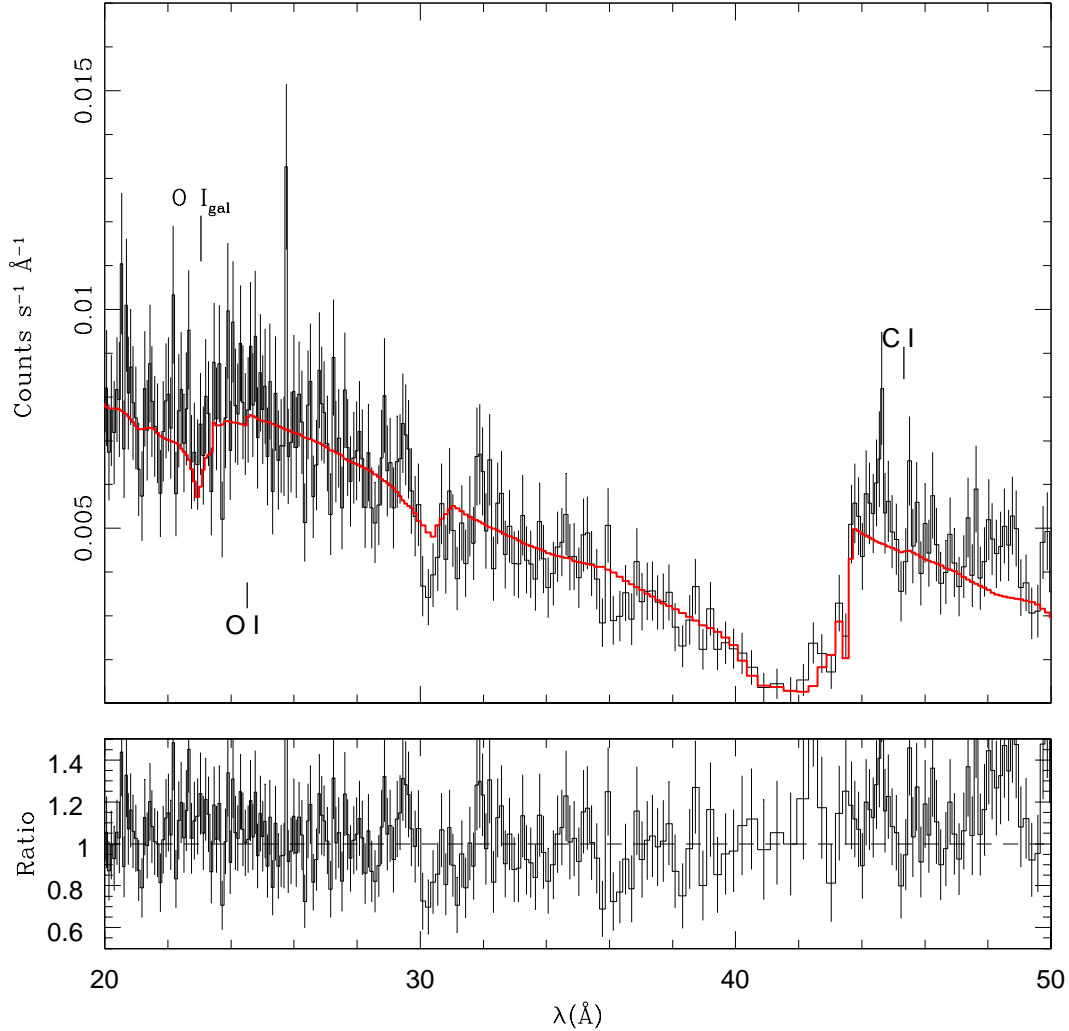}}}
\caption{
Test for the presence of a dusty warm absorber.
X-ray spectrum of \mr \space (observed frame), with a power-law model
modified by four edges (solid red line).
The 
edges corresponding to the H- and
He-like oxygen features are included in the model but they lie off the figure axis.
Marked are edges expected from dusty material
represented by bound-free transitions of neutral carbon and oxygen,
in the observer's frame. Also marked is the oxygen edge coming from the Galactic
absorption toward \mr.
\label{dusty1}}
\end{figure}

Dusty warm absorbers have been found to imprint
their hallmark on the X-ray spectra of several AGN 
\citep[e.g.][]{komossa1997a,komossa1998a,lee2001a}.
We investigate the possibility of the presence
of a dusty warm absorber by looking at the position of
two important edges, predicted to be strong and
noticeable in the X-ray band; the neutral carbon edge at 291 eV
(from the graphite species), and the O~{\sc i}
edge at 538 eV (from silicates).
Figure \ref{dusty1} shows the position of these
two edges
(the edges corresponding to the H- and
He-like oxygen features are included in the model but they lie off the figure axis),
in the observer's frame (also marked is
the Galactic neutral oxygen as O I$_{\rm gal}$).
These edges are not clearly seen in the spectrum,
but this does not preclude their potential addition 
and improvement to the model in the fitting.
In Table \ref{tbl4} we present the result of fitting
the spectrum of \mr \space with a model that
is composed of a Galactic absorbed power-law
modified by four edges, the K-edges of
C~{\sc i}, O~{\sc i}, O~{\sc vii} and O~{\sc viii}.
In order to search for fine features, we used a larger 
number of bins in the fit, 50 counts from the source per bin.
In terms of goodness of the fit, this is equal to
the {\tt pl $\times$ 2 edges} model,
with $\chi^2_{\nu}=1.22$ (for 587 d.o.f).
However, letting the four edges vary freely,
the C~{\sc i} optical depths has no lower limits.
An upper limit can be set on the dusty material if it exists.
We have enforced the $\tau_{\rm C~I}$ to be
equal to 0.08 (the upper limit of this measurement),
letting all the other optical depths free to vary, and re-fit. 
We set this as an upper limit on the
$\tau_{\rm C~I}$ parameter, since the fit does not get
worse with respect to the previous one (see row 2 in Table \ref{tbl4}).
This translates in a C~{\sc i} column density
$N_{\rm C~I}\approx 2 \times 10^{19}$ \cmn.
Doing the same with the $\tau_{\rm O~I}$ (taking the upper limit of the best-fit),
we obtain no significant change with respect to the best-fit $\chi^2$
(see row 3), and we consider  $\tau_{\rm O~I}=0.09$
as the upper limit on $\tau_{\rm O~I}$, which translates in $N_{\rm O~I}\approx 9 \times 10^{19}$ \cmn
\space
\citep[we used the photoionization cross sections from][ at their respective energies]{morrison1983a}.
For typical Galactic gas/dust ratios, our observed upper limit is not sensitive enough to provide significant constraints on the dust in \mr.
\subsection{Mean Warm Absorber
\label{xstar_calc1}}
\begin{figure}
\rotatebox{0}{\resizebox{15cm}{!}
{\plotone{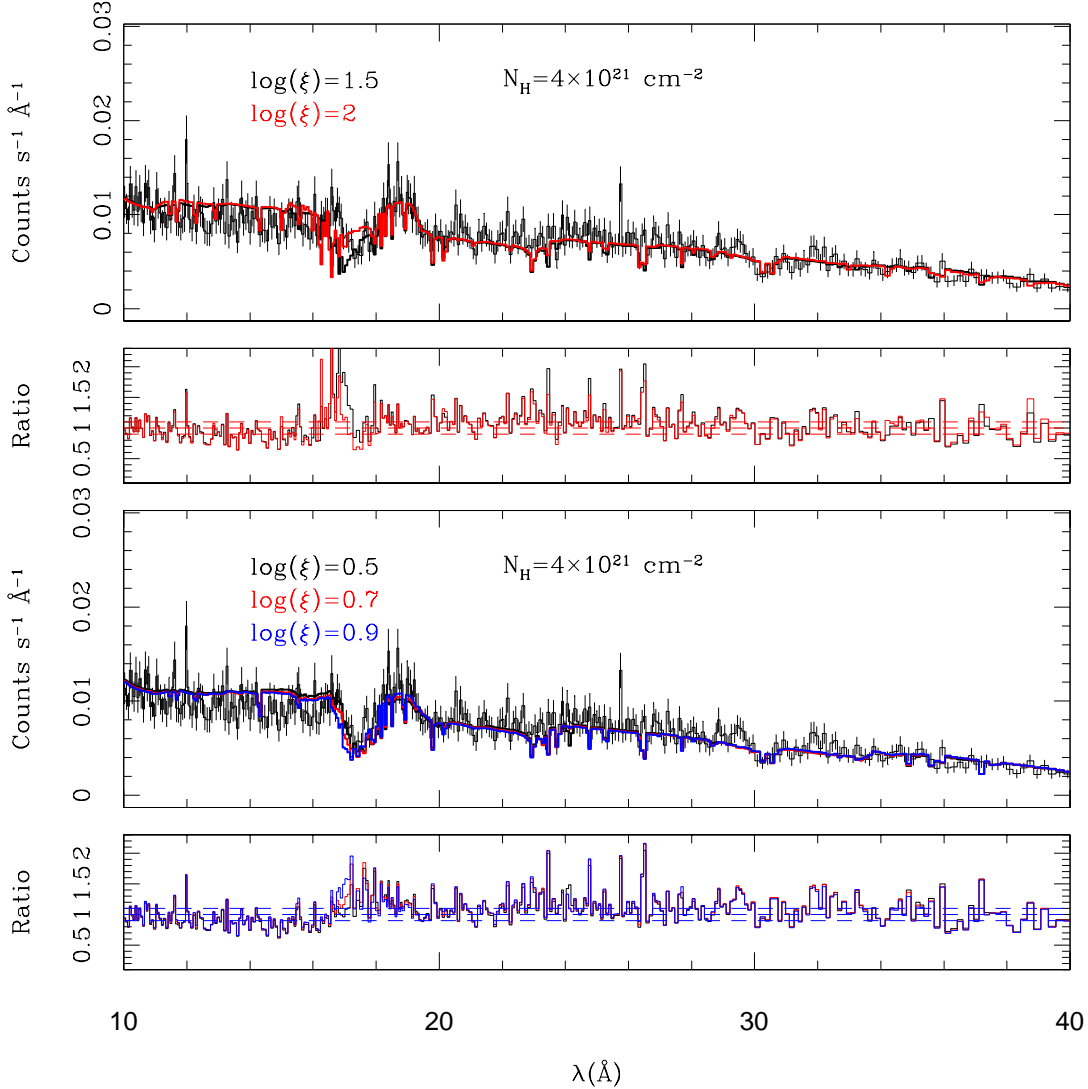}}}
\caption{
$10-40$ \AA \space spectrum of \mr \space and
XSTAR photoionization models
of different ionization parameters. 
We use $N_H=4 \times 10^{21}$ \cmn \sp for illustrative purposes.
{\bf Top panel}: High ionization state,
$\log(\xi)=1.5,2$; black and red lines respectively.
{\bf Bottom panel}: Lower ionization state,
$\log(\xi)=0.5,0.7,0.9$; black, red and blue lines respectively.
It can be observed that at higher $\log(\xi)$, the produced absorption
feature is at bluer wavelengths.
\label{utas1}}
\end{figure}

In this section we discuss the global view of the spectrum of \mr, in terms of a more
physical model. Visually, the strongest spectral feature is the broad absorption
structure around 17 ${\rm \AA}$ (see Figure \ref{utas1}), 
recognized as the hallmark of the warm
absorber \citep[e.g.,][]{george1998a}. Detailed calculations have demonstrated
that this structure is made (mainly) by the
O~{\sc vii}-K edge and by the Fe M-shell  $2p-3d$ unresolved transition array
(UTA) between 16$-$17 ${\rm \AA}$ \citep{behar2001b}. The ionization condition
under this structure is strong enough to dominate the emergent spectrum,
have been the subject of several theoretical and observational works 
\citep[e.g.,][]{krongold2003a,gu2006a},
due to in part, the particular sensitivity of this feature to the ionization
state of the gas responsible for the absorption.

To describe the state of the gas, we build a grid of photoionization models
assuming ionization and thermal equilibrium. Under conditions of ionization equilibrium
the state of the gas depends mostly  (apart from $n_{\rm H}$, abundances and column density) upon
the shape of the ionizing spectrum and the ionization parameter $\xi$,
that we define as in \cite{tarter1969a}:
\begin{equation}
\xi=\frac{4\pi F_{\rm ion}}{n_{\rm H}},
\label{xi}
\end{equation}
where $F_{\rm ion}$ is the total ionizing flux
($F_{\rm ion}=L_{\rm ion}/[4\pi r^2]$, see below for definition of $L_{\rm ion}$),
and $n_{\rm H}$ is the gas density.
We have carried out all the photoionization calculations using the
XSTAR
\footnote{Version 2.1kn6. UTAs included. However, this version does not include the
corrections (for some atomic line transitions) given by \cite{gu2006a}. 
We have re-computed our models based on an updated version of XSTAR (kindly provided by T. Kallman) which does
include the corrections of \cite{gu2006a}. We find that the change in the
outflow velocity is only $\sim 1$ \%. Part of the reason is that the LETGS has a spectral resolution
of $\approx 50$ m\AA, while the changes reported by \cite{gu2006a} were $\approx 15-45$ m\AA.}
code with the atomic data
of \cite{bautista2001a}. The code includes all the relevant atomic processes 
(including inner shell processes) and computes the emissivities and optical depths
of the most prominent X-ray and UV lines identified in AGN spectra.
Our models are based on spherical shells 
illuminated by a point-like X-ray continuum source.
The input parameters are the source spectrum, the gas
composition, the gas density $n_{\rm H}$, the column density and
the ionization parameter. The source spectrum is described by the spectral luminosity
$L_{\epsilon}=L_{\rm ion}f_{\epsilon}$, where $L_{\rm ion}$ is the integrated
luminosity from 1 to 1000 Ryd, and $\int_{1}^{\rm 1000~Ryd} f_{\epsilon}
d\epsilon=1$. 
The spectral function is taken to be the ionizing
spectrum of \cite{leighly2004a}. The gas consists
of the following elements, H, He, C, N, O, Ne, Mg, Si, S,
Ar, Ca and Fe. We use the abundances of \cite{grevesse1996a}
in all our models (we use the term {\it solar} for these abundances). 
We adopt a turbulent velocity of 100 \kms, and a gas density
$n_{\rm H}=10^{8}$ \cmd.

\subsubsection{Global Absorption
\label{abs1}}
We have fit the X-ray spectrum of \mr \space with our
XSTAR based photoionization models. First of all, we find that 
independent of column density, 
a relatively low ionization parameter is needed; else, the spectrum 
in the Fe UTA band is not well reproduced. We demonstrate this point in
Figure \ref{utas1}. It shows the resulting absorption
spectrum for several ionization states. 
For illustrative purposes, here we fix $N_{\rm H}=4\times 10^{21}$ \cmn.
In the top panel we can see the spectrum resulting at
high ionization parameter, $\log (\xi)=1.5$ and 2.
The bottom panel shows the
spectrum at lower ionization states with $\log (\xi)=0.5,~0.7$ and 0.9,
in black,  red and dark blue respectively.
From this we can rule out high ionization states
alone to describe this band \citep[though][ reported the use of a multi-component model to
represent the spectrum of \mr, we were not able to find a significant improvement on
the fit over the one-absorber component]{kaspi2004a},
because these models always
give large residuals in the UTA band (see below).

\begin{figure}
\rotatebox{0}{\resizebox{15cm}{!}
{\plotone{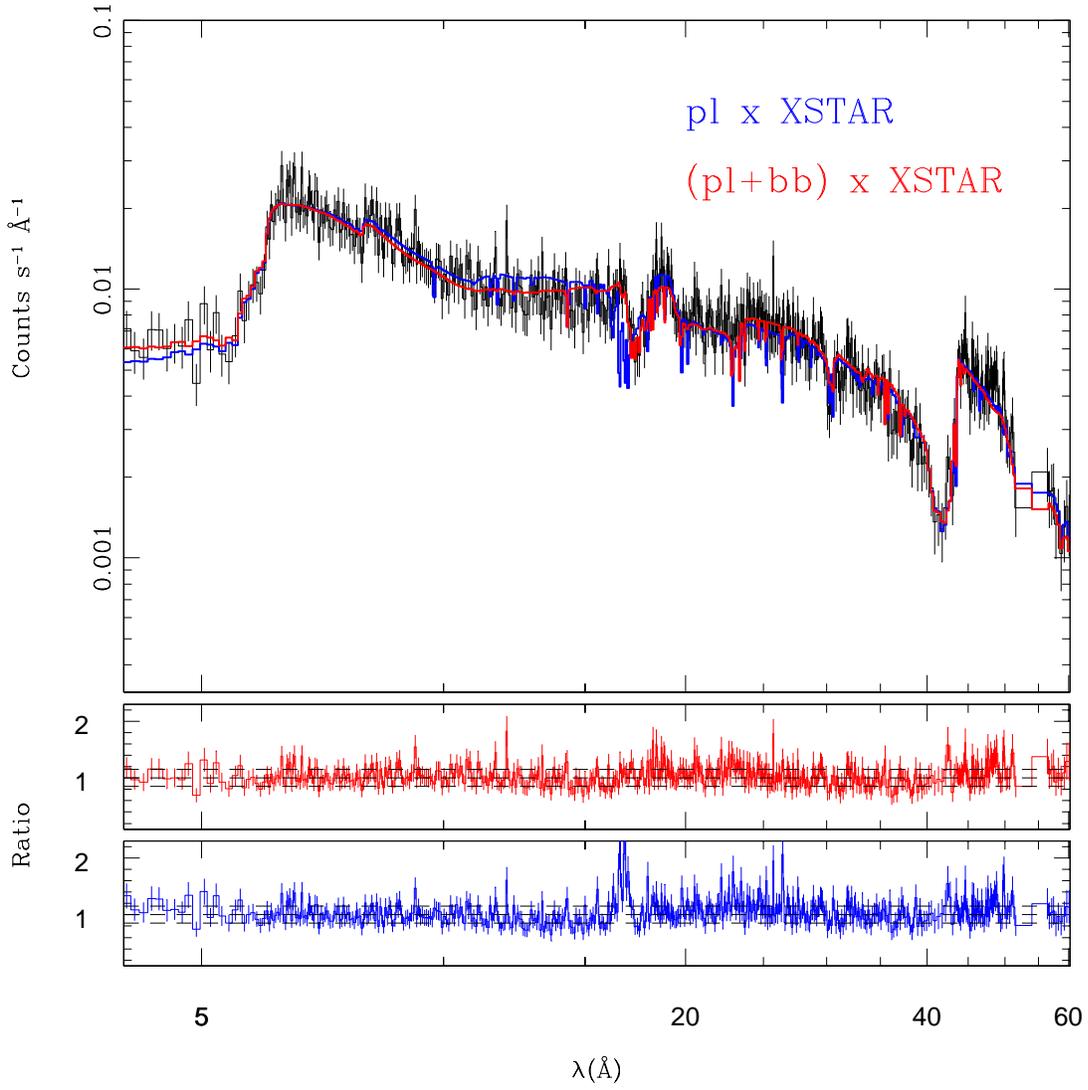}}}
\caption{
The best-fit XSTAR photoionization model
(blue). A residual is seen at $\sim 17 $ \AA. An important improvement
is achieved by the inclusion of a thermal component
(red). See Table \ref{tbl5} for parameter values.
\label{utas2}}
\end{figure}

Nevertheless, for our best-fit, single component photoionization model
({\tt pl $\times$ ${\rm xstar_{model}}\equiv {\rm X_1}$}, Figure \ref{utas2}),
we obtain the following parameters:
a column density of $N_{\rm H}=4.77 _{-0.29} ^{+0.31} \times 10^{21}$ \cmn,
an ionization parameter $\log(\xi)=1.83 _{-0.06} ^{+0.06}$ and
a photon index  $\Gamma_{\rm x}=1.96 _{-0.02} ^{+0.02}$. While the fit well describes
the long wavelength part of the absorption feature (with respect to the centroid
around $\sim$ 17.4 ${\rm \AA}$), there exists a large residual
in the shorter wavelength range ($\gtrsim$ 50 \%), resulting in
a fit with $\chi ^2 _{\nu}=1.54$ (for 587 d.o.f).
Motivated by the good
result of including a thermal component to describe any soft
excess, and also by the fact that the X-ray spectrum of this
object has been fitted before including a thermal component
\citep[for example in][]{gibson2005a}, we include in our model
a blackbody, characterized by its temperature
$kT$ (i.e., {\tt (pl+bb) $\times$ ${\rm xstar_{model}}\equiv {\rm X_2}$}).
The inclusion of the blackbody improved the fit, which gives
$\chi ^2 _{\nu}=1.27$ (and d.o.f $=585$). But we went further, and
inspected the possibility that the absorbing gas in \mr \space is
flowing outward from the center of the system with velocities
$\sim 0-2000$ \kms. For that purpose we shift the modeled spectrum
(the absorber)
by a grid of velocities keeping fixed the outflow velocity and letting
the others parameters of interest free to vary. For ${\rm X_1}$ we found
that it favors an outflow velocity $\approx 200$ \kms. In the
case of ${\rm X_2}$, the model points to an outflow velocity
$\approx 1200$ \kms.
After finding statistically motivated outflow
velocities for our models we were able of thawing them and find the
best-fit outflow velocity.

\begin{figure}
\rotatebox{0}{\resizebox{15cm}{!}
{\plotone{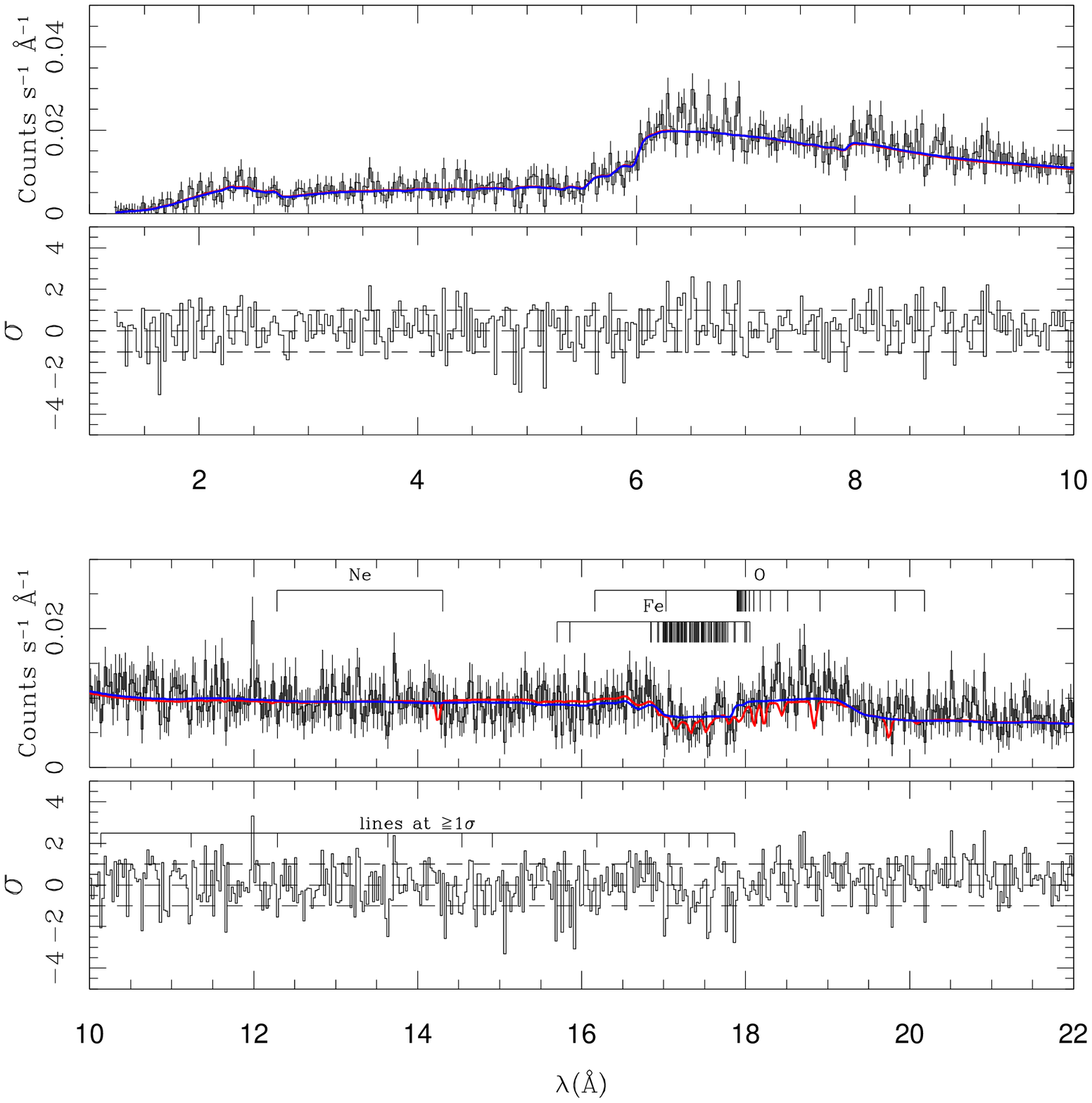}}}
\caption{
Comparison of the spectrum of \mr \sp with our 
best-fit XSTAR photoionization model which is presented in high resolution
(bin size 25 m\AA). The physical model is shifted by $-1100$ \kms. Predicted lines
of Ne, Fe, O, C and N are marked as labels in the top of the spectrum.
\label{utas3}}
\end{figure}

The {\tt ${\rm xstar_{model}}$} plus the blackbody (${\rm X_4}$) gives a
$\chi ^2 _{\nu}=1.22$ (with d.o.f $=584$), and the following best-fit
parameters: $N_{\rm H}=2.25 _{-0.25} ^{+0.28} \times 10^{21}$ \cmn,
$\log(\xi)=0.63 _{-0.06} ^{+0.06}$, $kT=89.5 _{-2.5} ^{+2.5}$ eV,
and $v_{\rm out}=-1100 _{-210} ^{+60}$ \kms.
An inspection of Figure \ref{utas2} reveals a good agreement
between the data and the model,
and the removal of the residual around $\sim$ 17 ${\rm \AA}$
(see ratio of data/model)
\footnote{
In order to see
if a multicomponent warm absorber system is supported by the data 
we repeat the above procedure adding a second warm absorber system to the
models, without (${\rm X_5}$) and with the blackbody (${\rm X_6}$).
In general the data do not well constrain the parameters of the second component.
And in any case, no significant improvement is seen by adding a second component to our model
(i.e, $\chi^2\approx 707$, $\Delta \chi^2\approx 6$ better than ${\rm X_4}$), and we do not
discuss this model further.
}.
Table \ref{tbl5} summarizes the parameter values of the model with and without
the thermal component, at rest (i.e., 0 \kms) and with the best-fit outflow velocity.
We present in Figure \ref{utas3} a high resolution (bin size 25 m\AA) version of
our best-photoionization model plotted over the spectrum of \mr \sp
(in the most interesting range to search for spectral lines,
$\sim 1-40$ \AA).
Here we fix the velocity outflow to the best-fit value found above (i.e., $-$1100 \kms) and allowed the others parameters
free to vary (i.e., $\Gamma_{\rm x}=1.64 _{-0.02} ^{+0.03}$, $N_{\rm H}=1.82 _{-0.22} ^{+0.26} \times 10^{21}$ \cmn,
$\log(\xi)=0.57 _{-0.13} ^{+0.09}$, $kT=95.2 _{-3.2} ^{+3.3}$ eV). 
We use our best-fit featureless continuum
(but including bound-free transitions)
as the underlying continuum for the search for possible spectral lines (also plotted in Figure \ref{utas3}), 
discussed in the next section.

\subsubsection{Absorption lines
\label{absor_lines1}}

There are several coincidences between the
theoretical prediction about where a line is located, and deviations of the data from the fit continuum
(the XSTAR featureless global continuum described above).
In order to seek for narrow absorption features (in a systematic way) in the spectrum, we use a finer
binned spectrum (bin size 25 m\AA). We use the following detection criteria:

\begin{enumerate}
\item The counts present in the data must deviate from the fit continuum by at least $1\sigma$.
As underlying continuum, we use the best-fit global featureless continuum, described above.
We use a global continuum fit because in that way we are less sensitive to local noise in the spectrum
\footnote{As a test of the robustness of our results, we have also determined the continuum locally around the line.
The continuum was fit locally ($\pm 0.5$ \AA) as a power-law. We find that all four key line candidates are still
present when a local continuum is used. To give an example: the equivalent width of the K$_{\beta}$ line of
N~{\sc vi} (see Table \ref{tbl7}) changes from $55 _{- 30} ^{+ 90}$ m\AA \sp to $\approx 70$ m\AA, when using a local continuum.
The measurements of the other three lines are unaffected by the local continuum fit.}.
\item The FWHM (within errors) of the line should not be significantly smaller than the instrumental FWHM, at the
correspondent wavelength.
\end{enumerate}

Not as a criterion, but as an extra piece of statistical evidence for the presence of lines, 
we use the maximum likelihood ratio test (MLR) to compare the  continuum model with the continuum+line model, 
and compute the line significance (see Table \ref{tbl6} column 4). $P(\geq T)$ is the probability that one would select 
the more complex model (continuum+line) when in fact the null hypothesis is correct (continuum alone).
We set a lower threshold of $P(\geq T) < 0.01$ (not accounting for number of trials).  
In fact, the four features which passed criteria 1 and 2 also had $P(\geq T)$ below this threshold
\citep[see][ for justification of using the MLR test]{band1997a}.

A list of candidate absorption lines is given in Table \ref{tbl6}. We measure the center (column 1), and the width
(in terms of $\sigma$ [in \kms], column 2) of each spectral structure using the best-fit continuum described above.

\begin{figure}
\rotatebox{0}{\resizebox{15cm}{!}
{\plotone{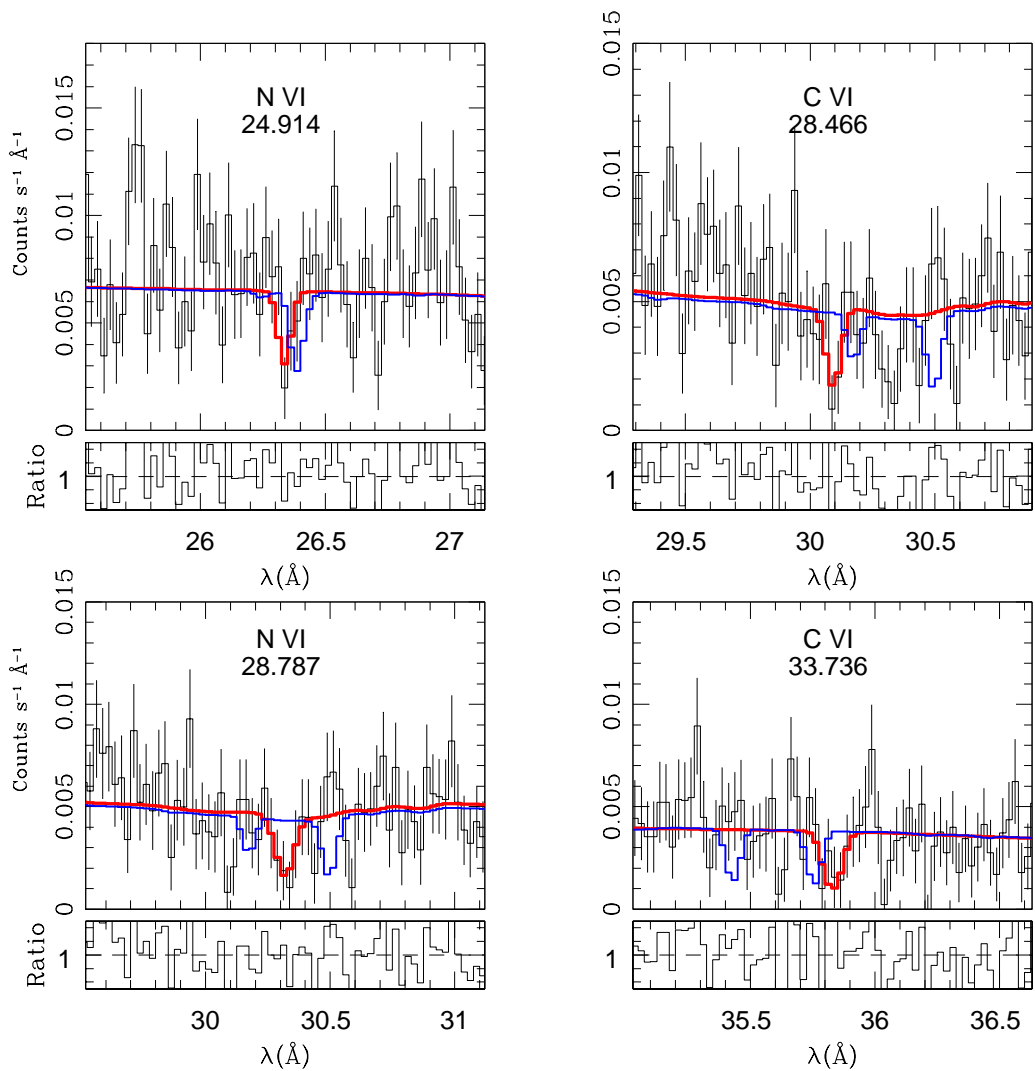}}}
\caption{
Strongest candidate absorption lines. The K$_{\alpha}$ and K$_{\beta}$ lines of
C~{\sc vi} and N~{\sc vi} show at least three outflow components at
$\sim -600$, $-2000$ and $-3000$ \kms. In red the Gaussian fit, in blue
the XSTAR model shifted by $-1100$ \kms.
\label{abs_lines1}}
\end{figure}

Figure \ref{abs_lines1} 
shows different portions of the spectrum containing absorption lines.
Among these lines, we only report measurements from four lines (Table \ref{tbl7}), located in a region where the
calibration of the LETGS is reliable and its sensitivity is high enough to report 90 \% confidence limits on their errors.
Deviations $\geq 1\sigma$, but have FWHM too low to be reported as individual
lines are still marked in the figure ($\sigma$-panel), but are not discussed here any further.

The measurements of the line parameters along with their identification is given in Table \ref{tbl7}. In the first
column we have the center of the line, and in the second column the equivalent width of the line (EW)
in milli-angstroms, computed with:
\begin{equation}
{\rm EW}=\int
\left[ 1-\frac{F_{\rm g}(\lambda)}{F_{\rm c}(\lambda)} \right]  d\lambda,
\label{ew}
\end{equation}
where $F_{\rm g}(\lambda)=
F_{\rm c}(\lambda)(1-\tau_0\exp[-\frac{(\lambda-\lambda_0)^2}{(2\sigma^2)}])$,
$F_{\rm c}(\lambda)$ is the continuum at the wavelength $\lambda$,
$\tau_0$ is the depth at the center of the line, $\sigma$ is the measured  width in units of \AA,
and $\lambda_0$ is the wavelength of the line in \AA \space (at the core).
The EW uncertainties are computed using the lower and upper limits of $\tau_0$ and $\sigma$
(i.e, ${\rm EW}_{\rm min}={\rm EW}[\tau_{\rm min},\sigma_{\rm min}]$ and 
${\rm EW}_{\rm max}={\rm EW}[\tau_{\rm max},\sigma_{\rm max}]$). 
In the fourth column we give the blueshift of the line (in \kms).
All these are resonance lines, product of electric dipole transitions
with the form $1s-np$ for C and $1s^2-1snp$ for N ions.

We discuss the issue of line identification in two step.
(1) As a first (simplifying) step we assume, that the spectrum is dominated by only one main warm absorber,
and that the column density and ionization parameter from the global fit to
the X-ray spectrum characterize this main warm absorber reasonably well.  Then, this best-fit global warm absorber model
guides us in identifying the absorption lines seen in the spectrum. In this first step, we only consider line
identification with transitions, which are actually being predicted to be strong (i.e., detectable) by this photoionization model. If no line feature
at zero velocity was predicted by the model, we allowed for a range of velocities (up to $\sim$ few thousand km/s - a range of outflow velocities
commonly observed in known warm absorbers). That procedure resulted in the line identification reported in Table \ref{tbl7}.
To give an example of a possible alternative
identification, we compare the strength of the line N~{\sc vi} at 24.9 \AA \sp (1s-2p, the one identified in Table \ref{tbl7})
with a potential alternative
identification, the line N~{\sc vii} at 24.8 \AA. The former line is much stronger than the latter:
The ratio of optical depth at the core is $\tau$(N~{\sc vi})/$\tau$(N~{\sc vii}) $\approx 8$,
for our best-fit global ionization parameter of $\log(\xi) = 0.6$. At this $\log(\xi)$,
there is almost 3.5 times more N$^{+5}$ than N$^{+6}$.

Assuming the one-absorber model is correct, the predicted strongest lines are the four line candidates presented in Table \ref{tbl7}.
However, we find that lines coming from the same ion are at different outflow velocity. This is a problem at the moment, and opens
up a second possibility, which we examined in a second step: (2) the possibility that we see several different warm absorber components
each with a different outflow velocity, and each characterized by a potentially different column density and ionization parameter.
The data quality, a single grating exposure of only 80 ks, 
does {\em not} allow to fit such a four-component absorber with so many free parameters. Therefore in step (2) we made the (simplifying) assumption that
the four identified lines are all at the same velocity.
We therefore assumed that all four lines are at zero velocity or at velocities up to $\approx 500$ \kms, in particular.
Independent of any warm absorber model (i.e., any specific $\xi$ and $N_H$), we searched line lists to find any consistent line identifications
for the four lines, fixing their wavelengths to the
quasar rest-frame (i.e., assuming zero velocity), or allowing velocities up to $\approx 500$ \kms. In that case, no line counterparts can be identified.

We therefore continue with solution (1), and discuss some implications of this solution.
Since most warm absorbers are not at rest, in order to identify lines, we allowed for a range of
outflow velocities, from $0-5000$ \kms.
We find that not all of the absorption line candidates are at the same outflow velocity, but we preliminary
identify 3 velocity systems
at $\sim $ $-600$ \kms, $-2000$ \kms, and $\sim -3,000$ \kms.
Further discussion of these three components, and a comparison 
with UV observations is provided in
\S \ref{discuss11}.

\section{Variability in the band ($0.6-1$) keV
\label{temp11}}

The region $\sim (0.6-1$) keV in the spectrum of \mr \space
is complex, due to the presence and blending
of three features: the Fe UTA, and the K-edges of oxygen
O~{\sc vii} and O~{\sc viii}. As a first step, we investigate temporal changes
assuming the spectrum is actually dominated by
the O~{\sc vii}  and O~{\sc viii} edges. Later in \S \ref{discu-time}, 
we carefully discuss the variability of the spectrum based on a photoionization
model, which includes a proper treatment of these features
including the UTA self-consistently. 
First, we split the observation in four (equally distributed, $\approx 11$ ks
of GTI) time intervals (1-4) and applied a {\tt pl $\times$ 2-edges} model to each set
of data. By fitting this model
(with the edges energy fixed at the values of O~{\sc vii} and O~{\sc viii}
K-edges, 739 and 871 eV respectively), we are able to notice changes
(in time) around the bound-free features, at $\sim$ 0.69 keV for
O~{\sc vii} and 0.81 keV for O~{\sc viii} in the observed frame
which in turn modify the parameter values of the model (see Table \ref{tbl8}).
To quantify these changes we looked at the parameters of the model
at each time interval. Due to the low S/N of the data these measurements have large errors.
Nevertheless,
one real inconsistency is seen in the
depth of the O~{\sc viii} edge ($\tau_{\rm OVIII}$) from time interval
(1) to (3): approximately
a factor of 3.4 in $\tau_{\rm OVIII}$ leads to a change in the count rate level of
$\approx 20$ \%, in the vicinity of the edge. Almost no change is seen in
$\tau_{\rm OVII}$, being consistently about the mean best-fit value 
found before; $\tau_{\rm OVII}\approx 0.6$.


We caution that the edge model is a too simple representation
of the spectrum in the ($0.6-1$) keV range,
where also the UTAs are located.
Therefore, we do not draw any physical implications
from it.   We discuss a more physical photoionization
model in \S \ref{discu-time}.


\section{Discussion
\label{discuss11}}

In this section we discuss the physical implications of our measurements,
which allow us  to draw conclusions about the kinematics
of the outflowing gas, ionization state of the warm absorber, and variability
of the absorption structure around
the oxygen edges seen in the spectrum of \mr.

\subsection{Identification of absorption lines} 

Identification of absorption lines in the X-ray spectra of AGN is important
because these lines carry a wealth of information on the physical
conditions in the ionized gas, and its link to other components of
the active nucleus.  

We have systematically searched for absorption and emission line
features in the X-ray spectrum of \mr, and have identified four absorption line
candidates corresponding to transitions in the ions C~{\sc vi} and N~{\sc vi}. 
These lines, if real, would imply a rather complex velocity field of the
absorber, with lines from the same ion indicating different outflow
velocities. In particular, it is of our concern the lack of corresponding
absorption from the element Oxygen{\footnote{we have carefully searched 
at the location of O~{\sc vii} at
$\lambda21.8$ \AA, and cannot identify
any feature.  This is consistent with \cite{gibson2005a}
while \cite{kaspi2004a} report some Oxygen
absorption features at $\sim 16-22$ \AA \sp in the RGS spectrum of
\mr \sp (their Table 3), albeit with low significance.}}.    
In case of solar abundances of the ionized gas, Oxygen absorption features are expected and 
have been detected in several AGN with warm absorbers 
\citep[e.g.,][]{kaastra2000a,sako2001a,lee2001a,blustin2003a,netzer2003a,fields2007a,costantini2007a,krongold2007a}
even though few of them with well-measured column densities.
On the other hand, it has also been found
that warm absorbers are complex, stratified, multi-component,
time-variable, 
and possibly of non-solar abundances, and the same may well hold for \mr.
With these cautious comments in mind, we briefly discuss implications 
of the line candidates we find. Ultimately, deeper grating observations
are needed to study the absorption line features and their temporal evolution.  

Independent information on the presence and velocity fields of 
ionized absorption components comes from a comparison of X-ray with with UV
observations.  
Firstly, in our LETG spectrum we identified a candidate gas component outflowing at
a velocity $v$ $\sim -600$ \kms, as indicated by the C~{\sc vi} $\lambda$33.736
line. This is consistent with UV measurements.
The UV spectrum of \mr \sp provides evidence for an absorbing system
at an outflow velocity of $v = -580$ \kms \sp \citep{kaspi2004a}. 
This component is seen through the
O~{\sc vi} $\lambda$1032,1038 and Ly$_{\alpha}$ lines, observed with the
STIS on board HST. Also (and reported by the same authors),
this component is in accordance with the velocity of $\approx -600$ \kms,
reflected by the Ly$_{\beta}$ line, observed with FUSE.
This is interesting, because it confirms the scenario where UV and X-ray absorber systems could be
physically related \citep{mathur1998a}. However, the centroid of 
our X-ray line candidate is measured with an
uncertainty that also allows to relate this X-ray absorption system with the 
average outflow velocity
of $\sim -300$ \kms \space reported by
\cite{monier2001a} using the Ly$_{\alpha}$, N~{\sc v}~$\lambda 1240$,
and C~{\sc iv}~$\lambda 1549$ lines in the ultraviolet (UV) band
(with the {\it HST-Faint Object Spectrograph} observation of
\mr).

For the computation of the mass-outflow rate, we assume that the
outflowing material in \mr \sp forms a spherical shell ($n\sim r^2$)
expanding at velocity $v$. Using the approximation $N\sim nr$,
we can write:
\begin{equation}
\dot{M}_{\rm out}=
4\pi \mu_H N r vf_{\rm cov},
\end{equation}
or
\begin{equation}
\dot{M}_{\rm out}\approx
0.048 N_{21} r_{18} \left( \frac{v}{1000 {\rm km~s^{-1}}}\right) f_{\rm cov}
~M_{\odot}~{\rm yr}^{-1},
\end{equation}
where $\mu_H$ is the mean mass per H atom (equal to $\frac{m_p}{0.7}$),
$N_{21}$ is the column density of the material in
units of $10^{21}$ \cmn, $r_{18}$  is the distance of the gas in units
of $10^{18}$  cm, and $f_{\rm cov}$ is the covering factor
($\Omega/[4\pi]$). Using our best-fit $N_{\rm H}\sim 10^{21.35}$,
an outflow velocity $v=1100$ \kms, at a
distance of $1.5 \times 10^{18}$ cm (distance chosen for illustrative purposes)
we have a mass-loss rate
$\dot{M}_{\rm out}\approx 0.18f_{\rm cov}$ $M_{\odot}~{\rm yr}^{-1}$.

In Table \ref{tbl9} is quoted our estimate of the
mass-loss rate of the LETG X-ray warm absorber system compared with previous values.
To compute the mass-loss rate, \cite{gibson2005a} used a high-velocity
outflow $v\sim 13,000$ \kms, concluding that $\dot{M}_{\rm out}$
could significantly
exceed the accretion rate ($\sim 0.2$ $M_{\odot}~{\rm yr}^{-1}$), by
$\sim 1-2$ orders
of magnitude, even taking the Fe~{\sc xxvi} line (see Table 7 of that paper).
On the other hand, \cite{monier2001a}, using an outflow velocity of 300 \kms,
estimate a mass-loss rate of $\sim$
0.9 $M_{\odot}~{\rm yr}^{-1}$ for
a 10\% covering factor, putting this value close to the accretion
rate.
Our computation of  $M_{\odot}$ is more in agreement with the value found by
these last authors.
The kinematic energy rate carried away by the flow spans
$\approx (0.07-1) \times 10^{42}f_{\rm cov}$ \ergs.

\subsection{Absorption variability during the observation 
\label{discu-time}}

Warm absorbers are known to vary on short and long time scales. 
We find indications for variability during
the LETG observation of \mr \sp
in the wavelength
range that includes the oxygen absorption edges
and the Fe UTA features.
Both, the UTAs  \cite[e.g.,][]{krongold2005b}
and/or the absorption edges could be variable. 
Mechanisms of variability include changes of the ionization 
state of the absorber to changes in the ionizing continuum,
changes of the internal structure of the absorber, and
clouds crossing our line of sight.

\begin{figure}
\rotatebox{0}{\resizebox{15cm}{!}
{\plotone{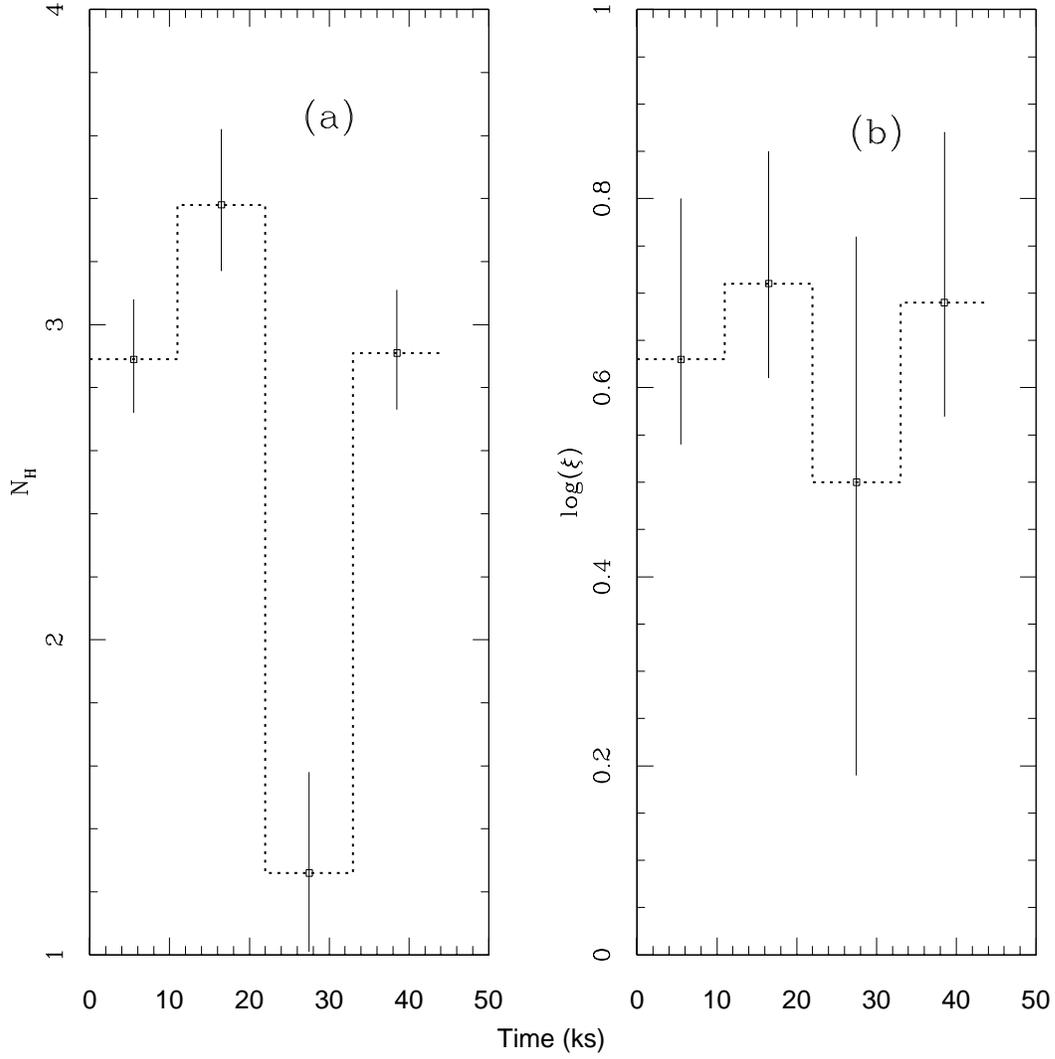}}}
\caption{
Time variability of the parameters for our model $X_4$ with
the column density
$N_{\rm H}$ of the warm absorber free to vary. (a) Column density in units
of $10^{21}$ \cmn. (b) Ionization parameter $\log(\xi)$, with $\xi$ 
in units of erg cm s$^{-1}$.
\label{xstar_N1}}
\end{figure}

\begin{figure}
\rotatebox{0}{\resizebox{15cm}{!}
{\plotone{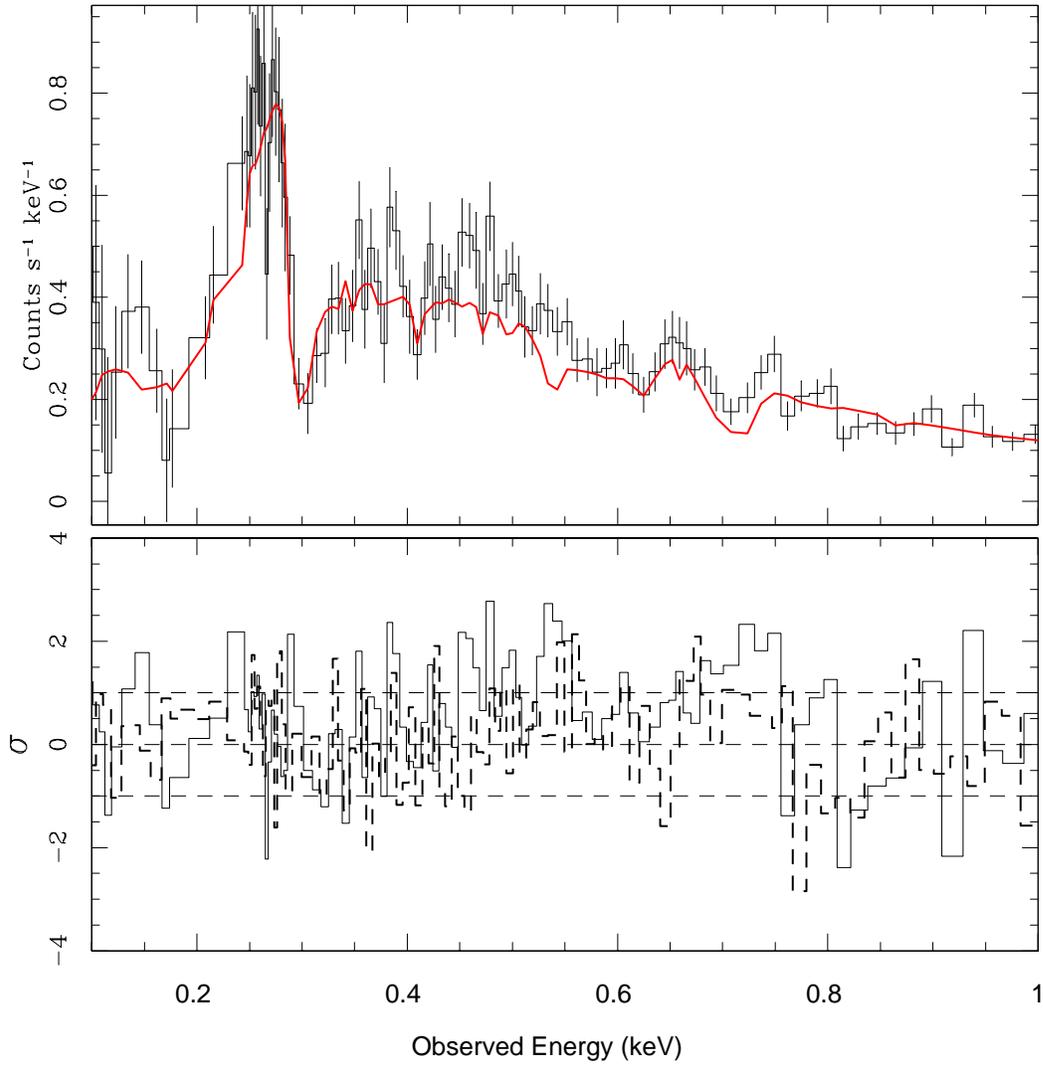}}}
\caption{Best-fit model ($X_4$) resulting from epoch (1) with $\chi_{\nu} ^2=1.1$ (interrupted-line),
applied to the data of epoch (3) (solid-line). Deviations are seen in the soft X-ray band ($\chi_{\nu} ^2=1.7$).
\label{sigma_ratio}}
\end{figure}

In order to find clues on the variability mechanism,
we have fit our most successful warm absorber photoionization model
(model  $X_4$; i.e., {\tt (pl+bb) $\times $ xstar}, with $v_{\rm out}=-1100$ \kms) 
separately to the four subsets of the total observation (Fig. \ref{light1}).  
We have explored the effects of variable column density, ionization parameter, and 
intensity of the continuum (black body and powerlaw) spectral components. 
The strongest effect is a change in column density of the ionizing material in epoch (3);
formally requiring (at $\gtrsim 6 \sigma$) a smaller column density in order to fit successfully
that epoch (Fig. \ref{xstar_N1}). In order to re-check whether the data
do indicate variability between the different epochs,
we have used the best-fit model of epoch (1) and compared
it to the data of epoch (3). In the comparison, we have fixed
all the parameters (of our model $X_4$) to those derived for epoch (1), and then inspected
the residuals during epoch (3). Clear deviations are seen
(Figure \ref{sigma_ratio}), demonstrating independently the presence of spectral
changes throughout the observation.

Changes in the luminosity would affect the ionization state of
the absorber, but the average countrate or MR2251-178 does not vary
throughout the observation. 
Taken at face value, a true change in column
density would require a change in density and/or thickness of the absorber,
and we first briefly comment on this possibility.
Assuming the thickness of the absorber has not changed in such a short time, 
the density of the absorber would need to change by a factor of 3 to account for the change 
in $N_H$ between periods (1) and (3).
However, changing the density of the whole extended
gas (or a fraction of the gas by a large amount),
is not easy in such a short time scale.
Re-arrangements of clouds crossing our line-of-sight would have to occur within $\sim$ 10 ks, which is very unlikely.
Alternatively we may
see non-equilibrium ionization
effects in the gas, in which the material responds 
with some time delay in a complicated way to previous variability in the ionizing
continuum, which was not within our observation time window.
Deeper observations with longer time bases are required to investigate these possibilities
further.

\section{Conclusions
\label{conclu11}}

The measured ($0.1-2$) keV flux
$f_{(0.1-2~{\rm keV})}=2.58 _{-0.03} ^{+0.04}\times 10^{-11}$ \flux \sp
implies that \mr \space is in
a relatively low state.
The soft X-ray luminosity
amounts to $L_{(0.1-2~{\rm keV})}=2.54\times 10^{44}$ \ergs.

We did not find any strong evidence for the presence
of an extra cold material with column density of $N_{\rm H}\sim 10^{21}$ \cmn \sp
toward \mr \sp (if part of this is located along our line of sight).
Based on different spectral fits,
we set the upper limit of this component
to $N_{\rm cold} \approx 1.2 \times 10^{20}$ \cmn.

As for previous observations of \mr, power-law plus blackbody
does not provide a successful fit to the X-ray spectrum.
The addition of a warm absorber improves the fit significatively.

Based on XSTAR photoionization modelling,
we find the observation
to be consistent with ionized
absorbing material in our line of sight,
column density
$N_{\rm H}\approx 2 \times 10^{21}$ \cmn, and an ionization parameter of 
$\log (\xi) \approx 0.6$. The inclusion of an additional thermal
component  is important because otherwise:
(1) the absorption structure around 18 \AA \sp is not well reproduced,
(2) the ionization state of the gas would be higher, bringing a lot of problems
in the identification of the atomic line transition. The temperature
of the blackbody used to represent this thermal component
is $kT \approx 90$ eV.

We find four line candidates
at moderate confidence level.
If these lines are real, their presence would imply
three components
traveling at velocities $\sim 600$, 2000 and 3000 \kms.
We compute $\dot{M}\sim 0.01-0.1 M_{\odot}~{\rm yr}^{-1}$
and a kinematic energy of
$\sim 10^{40-42}$ \ergs,
using the C~{\sc vi} and N~{\sc vi}
lines outflowing at velocities of
$\sim 2000-3000$ \kms.
Using $f_{\rm cov}=0.1$, this value is a factor of $\sim 20$ less than the accretion rate of the system.
However, we caution that these candidates need to be confirmed
with future observations. The lack of corresponding oxygen absorption is a problem
at present.

We do not find positive evidence for a dusty warm absorber.
The spectrum of \mr \space allows us to set an upper limit
of $N_{\mbox{\tiny C I}}\approx 2 \times 10^{19}$ \cmn,
and $N_{\mbox{\tiny O I}}\approx 9\times 10^{19}$ \cmn,
if dusty material is present in the nucleus of
\mr.

We find changes in the absorption structure
of \mr \space during the observation, not accompanied by changes
in the observed continuum luminosity.
Possibly, we see non-equilibrium effects in the ionization of the gas
responding to previous changes in luminosity, or changes in density;
the true mechanism can only be uncovered with deeper follow-up observations.

\acknowledgments

We thank the anonymous referee for many constructive comments.
The observation would not have been
possible to analyze without the enormous
effort of the {\it Chandra} team.

\clearpage
\begin{deluxetable}{crrr}
\tabletypesize{\scriptsize}
\tablecaption{{\it Chandra} Observation Log of \mr.
\label{tbl1}}
\tablewidth{0pt}
\tablehead{
\colhead{Sequence Number} & \colhead{UT Start}   & \colhead{UT End}   &
\colhead{Time (ks)}
}
\startdata
700405 &        2002 Dec 23, 16:35 &    2002 Dec 24, 14:54 &     78.5 \\
\enddata
\end{deluxetable}
\begin{deluxetable}{rrrrrrrr}
\tablecolumns{8}
\tablewidth{0pc}
\tablecaption{Cold absorption fit results
\label{tbl2}}
\tablehead{
\colhead{$N_{\rm H}$ $^{\rm a}$}&
\colhead{$\Gamma_{\rm x}$}&
\colhead{Norm$^{\rm h}$}&
\colhead{$\Gamma_{\rm x}(2)$}&
\colhead{Norm$^{\rm h}(2)$}&
\colhead{$kT$ $^{\rm i}$}&
\colhead{Norm $^{\rm j}$}&
\colhead{$\chi ^2 _{\nu}$/(d.o.f)}}
\startdata
$1.68 _{-0.07} ^{+0.07}$ $^{\rm b}$& $1.79 _{-0.02} ^{+0.02}$ & $5.34 _{-0.06} ^{+0.06} $  &\nodata & \nodata & \nodata & \nodata& 2.55/306\\
$2.77 $ $^{\rm c}$& $1.94 _{-0.02} ^{+0.02}$ & $5.65 _{-0.06} ^{+0.06} $&\nodata & \nodata &\nodata & \nodata& 3.00/307\\
$3.27 $ $^{\rm d}$& $2.00 _{-0.02} ^{+0.02}$ & $5.78 _{-0.06} ^{+0.06} $&\nodata & \nodata &\nodata & \nodata& 3.40/307\\
$3.77 $ $^{\rm e}$& $2.05 _{-0.02} ^{+0.02}$ & $5.91 _{-0.06} ^{+0.06} $&\nodata & \nodata &\nodata & \nodata& 3.85/307\\
$ \lesssim 10^{-6} $ $^{\rm b}$& $1.52 _{-0.02} ^{+0.02}$ & $4.70 _{-0.05} ^{+0.07}$ & \nodata & \nodata & $80.3 _{-1.3} ^{+3.0}$ & $1.04 _{-0.05} ^{+0.03}$ & 1.64/304\\
$1.24  _{-0.09} ^{+0.10}$ $^{\rm b}$& $3.47 _{-0.03} ^{+0.03}$ & $1.20 _{-0.03} ^{+0.03}$ & $1.41 _{-0.03} ^{+0.03}$ & $3.97 _{-0.06} ^{+0.06}$ &\nodata & \nodata & 1.81/304\\
$1.74$ $^{\rm f}$ & $3.70 _{-0.03} ^{+0.03}$ & $1.04 _{-0.03} ^{+0.03}$ & $1.44 _{-0.02} ^{+0.03}$ & $4.16 _{-0.06} ^{+0.06}$ & \nodata &\nodata & 1.81/305\\
$2.24$ $^{\rm g}$ & $3.94 _{-0.03} ^{+0.03}$ & $0.91 _{-0.02} ^{+0.02}$ & $1.47 _{-0.02} ^{+0.03}$ & $4.34 _{-0.06} ^{+0.06}$ & \nodata &\nodata& 1.84/305\\
\enddata
\tablecomments{
The error parameters are 90 \% confidence limits.
(a) Absorber column density in units of $10^{20}$ \cmn.
(b) Free parameter.
(c) Fixed to Galactic value.
(d) Fixed to Galactic value plus $\Delta N_{\rm H}$, where $\Delta N_{\rm H}=5 \times 10^{19}$ \cmn.
(e) Fixed to Galactic value plus 2 $\times \Delta N_{\rm H}$.
(f) Extra absorber column density
(shifted by $z=0.06398$, model {\tt zphabs} in XSPEC), fixed to the best value plus $\Delta N_{\rm H}$.
The fit is corrected by the Galactic absorption.
(g) Extra absorber column density
(shifted by $z=0.06398$, model {\tt zphabs} in XSPEC), fixed to the best value plus 2 $\times \Delta N_{\rm H}$.
The fit is corrected by the Galactic absorption.
(h) Powerlaw normalization, $\times 10^{-3}$ photons keV$^{-1}$cm$^{-2}$s$^{-1}$ at 1 keV.
(i) Temperature of the blackbody in eV.
(j) $\times 10^{-6}$ in units of $L_{39}/D^2_{10}$, where $L_{39}$ is the luminosity of 
the source in units of 10$^{39}$ ${\rm erg~s^{-1}}$,
and $D_{10}$ the distance to the source in units of 10 kpc.}
\end{deluxetable}
\clearpage
\begin{deluxetable}{clllllll}
\tablecolumns{8}
\tablewidth{0pc}
\tablecaption{Results from simple model fits
\label{tbl3}}
\tablehead{
\colhead{Model$^{\rm b}$}  & \colhead{$\Gamma_{\rm x}$}
& \colhead{Norm$^{\rm c}$}
& \colhead{$\tau_{{\rm O~VII}}$} & \colhead{$\tau_{{\rm O~VIII}}$}
& \colhead{kT$^{\rm d}$}& \colhead{Norm$^{\rm e}$}
& \colhead{$\chi ^2 _{\nu}$}/(d.o.f)}
\startdata
$[1]$ & $1.94 _{-0.02} ^{+0.02}$ & $5.65 _{-0.06} ^{+0.06}$& \nodata & \nodata & \nodata & \nodata & 3.00/(307)\\
$[2]$ & $1.53 _{-0.02} ^{+0.02}$ & $4.73 _{-0.06} ^{+0.06}$& \nodata & \nodata & $81.5 _{- 2.0} ^{+ 2.0}$ & $9.94 _{-0.38} ^{+0.38}$ & 1.63/(305)\\
$[3]$ & $1.92 _{-0.02} ^{+0.02}$ & $6.53 _{-0.07} ^{+0.07}$& $0.56 _{-0.04} ^{+0.05}$ & $0.22 _{-0.04} ^{+0.05} $ & \nodata & \nodata & 1.38/(305)\\
$[4]$ & $1.85 _{-0.02} ^{+0.02}$ & $6.35 _{-0.07} ^{+0.07}$& $0.51 _{-0.04} ^{+0.05}$ & $0.21 _{-0.04} ^{+0.05} $ & 
$44.1 _{- 3.0} ^{+ 3.0}$ & $5.57 _{-0.92} ^{+0.92} $ & 1.25/(303)\\
\enddata
\tablecomments{
All the models include absorption fixed at the Galactic value of
$N_{\rm H}=2.77 \times 10^{20}$ \cmn. The error parameters are 90 \% confidence limits.
(b): Model [1]: {\tt pl}; [2]: {\tt pl+bb}; [3]: {\tt pl$\times$2 edges}; [4]: {\tt (pl+bb)$\times$2 edges};
{\tt pl=powerlaw}; {\tt bb=blackbody}.
(c) Powerlaw normalization, in units of $10^{-3}$ photons keV$^{-1}$cm$^{-2}$s$^{-1}$ at 1 keV.
(d) Temperature of the blackbody in eV.
(e) $\times 10^{-5}$ in units of $L_{39}/D^2_{10}$, where $L_{39}$ is the luminosity of the source in units of 10$^{39}$ ${\rm erg~s^{-1}}$,
and $D_{10}$ the distance to the source in units of 10 kpc.}
\end{deluxetable}
\clearpage
\begin{deluxetable}{rrrrrrr}
\tablecolumns{7}
\tablewidth{0pc}
\tablecaption{Dusty warm absorber fit results
\label{tbl4}}
\tablehead{
\colhead{$\Gamma_{\rm x}$}&
\colhead{Norm$^{\rm b}$}&
\colhead{$\tau_{\rm O~VII}$}&
\colhead{$\tau_{\rm O~VIII}$}&
\colhead{$\tau_{\rm C~I}$}&
\colhead{$\tau_{\rm O~I}$}&
\colhead{$\chi ^2 _{\nu}$/(d.o.f)$^{\rm e}$}}
\startdata
$1.92 _{-0.02} ^{+0.02}$ & $6.53 _{-0.07} ^{+0.07}$
&$0.55 _{-0.04} ^{+0.05}$ & $0.23 _{-0.05} ^{+0.04} $
&$0.02 ^{+0.06}$ & $0.05 _{-0.04} ^{+0.04} $  & 1.22/585\\
$1.93 _{-0.02} ^{+0.02}$ & $6.58 _{-0.07} ^{+0.07}$
&$0.55 _{-0.04} ^{+0.05}$ & $0.23 _{-0.04} ^{+0.05} $
&0.06$^{\rm c}$ & $0.06 _{-0.04} ^{+0.04} $ & 1.22/586\\
$1.93 _{-0.02} ^{+0.02}$ & $6.58 _{-0.07} ^{+0.07} $
&$0.54 _{-0.04} ^{+0.05}$ & $0.24 _{-0.04} ^{+0.05} $
&$0.04 ^{+0.06}$ & 0.09$^{\rm d}$  & 1.22/586\\
\enddata
\tablecomments{
The error parameters are 90 \% confidence limits.
(b) Powerlaw normalization, $\times 10^{-3}$ photons keV$^{-1}$cm$^{-2}$s$^{-1}$ at 1 keV.
(c) Fitting, enforcing $\tau_{\rm C~I}=0.08$, and $\tau_{\rm O~I}$ free.
(d) Fitting, enforcing $\tau_{\rm O~I}=0.09$, and $\tau_{\rm C~I}$ free.
(e) In order to search for fine features, we used a larger number of bins in the fit,
50 counts from the source per bin.
For comparison the best fit model of the {\tt pl $\times$ 2edges} model (using this data)
gives $\chi_{\nu}^2=1.22$/(d.o.f=587).}
\end{deluxetable}
\clearpage
\begin{deluxetable}{lcccc}
\tablecolumns{5}
\tablewidth{0pt}
\tablecaption{Fit results for photoionization models.
\label{tbl5}}
\tablehead{
\colhead{Parameter}&
\colhead{${\rm X_1}$}&
\colhead{${\rm X_2}$}&
\colhead{${\rm X_3}$}&
\colhead{${\rm X_4}$}}
\startdata
$\Gamma_{\rm x}$ 		& $1.96_{-0.02} ^{+0.02}$  & $1.66 _{-0.02} ^{+0.02}$ 	&  $1.97 _{-0.02} ^{+0.02}$ &  $1.66 _{-0.02} ^{+0.02}$ \\
Norm $^{\rm a}$			& $6.84_{-0.07} ^{+0.07}$  & $5.42_{-0.07} ^{+0.07}$   	&  $6.82 _{-0.07} ^{+0.07}$ &  $5.38 _{-0.07} ^{+0.07}$ \\
$kT$ (eV)               	& \nodata                  & $89.9_{-2.5} ^{+2.6}$   	&  \nodata                  &  $89.5_{-2.5} ^{+2.5}$     \\
${\rm Norm}_{\tt bb}$ $^{\rm b}$& \nodata                  & $10.21_{-0.42} ^{+0.42}$   &  \nodata                  &  $10.29_{-0.40} ^{+0.44}$  \\
$N_{\rm H}$ $^{\rm c}$   		& $4.77_{-0.29} ^{+0.31}$  & $2.44 _{-0.29} ^{+0.13} $ 	&  $4.28 _{-0.21} ^{+0.29}$ &  $2.25_{-0.25} ^{+0.28}$   \\
$\log(\xi)$        	  	& $1.83_{-0.06} ^{+0.06}$  & $0.64_{-0.06} ^{+0.11}$	&  $1.72 _{-0.07} ^{+0.07}$ &  $0.63_{-0.06} ^{+0.06}$   \\
$v_{\rm out}$ $^{\rm d}$  	& 0                        & 0                          &  $-340 _{-20} ^{+120}$    &  $-1100_{-210} ^{+60}$     \\
$\chi^2_{\nu}$/(d.o.f)  	& 1.54/(587) 		   & 1.27/(585)     		&  1.52/(586)     	    &  1.22/(584)		\\
\enddata
\tablecomments{
(a) Powerlaw normalization, in units of $\times 10^{-3}$ photons keV$^{-1}$cm$^{-2}$s$^{-1}$ at 1 keV.
(b) $\times 10^{-5}$ in units of $L_{39}/D^2_{10}$, where $L_{39}$ is the luminosity of the source 
in units of 10$^{39}$ ${\rm erg~s^{-1}}$, and $D_{10}$ the distance to the source in units of 10 kpc.
(c) Column density of the ionized material in units of $\times 10^{21}$ \cmn.
(d) Outflow velocity of the ionized gas in units of \kms. Models ${\rm X_1}-{\rm X_4}$ are described
in the text (see section \ref{abs1}).
}
\end{deluxetable}
\clearpage
\begin{deluxetable}{llll}
\tablecolumns{4}
\tablewidth{0pc}
\tablecaption{Gaussian absorption line candidates
\label{tbl6}}
\tablehead{
\colhead{Obs. $\lambda$} &
\colhead{$\sigma_{\rm measured}$}&
\colhead{$\sigma_{\rm LETG}$} &
\colhead{MLR} \\
\colhead{(\AA)} & \colhead{(\kms)} &
\colhead{(\kms)} &
\colhead{$P(\ge T)$} }
\startdata
$26.34 _{-0.01} ^{+0.03}$ & $ 160 _{-  90} ^{+ 100}$ &  242  &  $7.38 \times 10^{-3}$\\
$30.09 _{-0.02} ^{+0.02}$ & $ 190 _{-  90} ^{+ 130}$ &  212  &  $1.13 \times 10^{-3}$\\
$30.32 _{-0.02} ^{+0.03}$ & $ 280 _{- 120} ^{+ 290}$ &  210  &  $7.06 \times 10^{-4}$\\
$35.83 _{-0.03} ^{+0.03}$ & $ 220 _{- 110} ^{+ 130}$ &  178  &  $1.17 \times 10^{-2}$\\
\enddata
\tablecomments{
Error parameters are 90 \% confidence limits.
The final significance is less than the one shown in column 4, since
the number of trials is not considered. 
The MLR test is described in Section \ref{absor_lines1}.
}
\end{deluxetable}
\clearpage
\begin{deluxetable}{rrrrr}
\tablecolumns{5}
\tablewidth{0pc}
\tablecaption{Identification of absorption line candidates
\label{tbl7}}
\tablehead{
\colhead{$\lambda_{\rm obs}$} &
\colhead{EW} &
\colhead{Ion and atomic}&
\colhead{$\lambda_{\rm lab}$}&
\colhead{v$_{\rm shift}$}\\
\colhead{(\AA)} &
\colhead{(m\AA)} &
\colhead{transition} &
\colhead{(\AA)} &
\colhead{(\kms)}}
\startdata
$26.34 _{-0.01} ^{+0.03}$ & $ 55 _{- 30} ^{+ 90}$ &  N~{\sc vi}  1s$^2$ $^1$S $-$ 1s3p$^1$P$^{\rm o}$ & 24.914 & $ -2030 _{-  120} ^{+ 360}$  \\
$30.09 _{-0.02} ^{+0.02}$ & $ 63 _{- 35} ^{+ 84}$ &  C~{\sc vi}  1s $^2$S $-$ 3p$^2$P$^{\rm o}$       & 28.466 & $ -2080 _{-  210} ^{+ 210}$  \\
$30.32 _{-0.02} ^{+0.03}$ & $ 70 _{- 40} ^{+ 76}$ &  N~{\sc vi}  1s$^2$ $^1$S $-$ 1s2p$^1$P$^{\rm o}$ & 28.787 & $ -3220 _{-  210} ^{+ 310}$  \\
$35.83 _{-0.03} ^{+0.03}$ & $ 86 _{- 30} ^{+ 98}$ &  C~{\sc vi}  1s $^2$S $-$ 2p$^2$P$^{\rm o}$       & 33.736 & $ - 580 _{-  270} ^{+ 270}$  \\
\enddata
\tablecomments{
Errors are 90\% confidence limits.
}
\end{deluxetable}
\clearpage
\begin{deluxetable}{rrrrrrrrr}
\rotate
\tablecolumns{9}
\tablewidth{0pc}
\tablecaption{Variability of the oxygen edges$^a$
\label{tbl8}}
\tablehead{
\colhead{Period}& 
\colhead{$t_{\rm start}^b$}&
\colhead{$t_{\rm end}^b$}&
\colhead{total}&
\colhead{$\Gamma_{\rm x}$}&
\colhead{Norm$^c$}&
\colhead{$\tau_{O^{+6}}$}& 
\colhead{$\tau_{O^{+7}}$}&
\colhead{$\chi ^2 _{\nu}$/(d.o.f)}\\
\colhead{}&
\colhead{$\times 10^8$ (s)}&
\colhead{$\times 10^8$ (s)}&
\colhead{counts ($\times 10^4$)}&
\colhead{}&
\colhead{}}
\startdata
(1)&1.57048747&1.570653992&1.8298&$1.90 _{-0.04} ^{+0.04}$ & $6.11 _{-0.13} ^{+0.13}$ &$0.57 _{-0.09} ^{+0.09}$ & $0.09 ^{+0.08}$ & 1.19/161 \\
(2)&1.570653993&1.570819941&1.6104&$1.89 _{-0.04} ^{+0.04}$ & $6.58 _{-0.14} ^{+0.14}$&$0.67 _{-0.09} ^{+0.10}$ & $0.25 _{-0.09} ^{+0.10} $ & 1.01/145\\
(3)&1.570819942&1.570982180&1.7391&$1.91 _{-0.04} ^{+0.04}$ & $6.62 _{-0.14} ^{+0.14}$&$0.46 _{-0.08} ^{+0.09}$ & $0.31 _{-0.09} ^{+0.09} $ & 1.32/159\\
(4)&1.570982181&1.571276092&1.5842&$1.96 _{-0.04} ^{+0.04}$ & $6.61 _{-0.15} ^{+0.15}$&$0.57 _{-0.09} ^{+0.10}$ & $0.22 _{-0.09} ^{+0.10} $& 1.39/144\\
\enddata
\tablenotetext{a}{The model includes Galactic absorption of
$N_{\rm H}=2.77 \times 10^{20}$ \cmn. 
The error parameters are 90 \% confidence limits
computed with the spectrum adaptively binned to have
at least 50 counts from the source per bin.}
\tablenotetext{b}{In the detector (original) time. These intervals exclude data from the ``bad time intervals" (section \ref{lc1})}
\tablenotetext{c}{Powerlaw normalization, in units of $10^{-3}$ photons keV$^{-1}$cm$^{-2}$s$^{-1}$ at 1 keV.}
\end{deluxetable}
\clearpage
\begin{deluxetable}{rrrrrr}
\tablecolumns{6}
\tablewidth{0pc}
\tablecaption{Outflow masses and kinematic energy rates
\label{tbl9}}
\tablehead{
\colhead{Ion$^a$}&
\colhead{$\dot{M}$\citep{monier2001a}$^b$}&
\colhead{$\dot{M}$$^{\rm e}$}&
\colhead{$\frac{1}{2}\dot{M}v^2$$^{\rm e}$}&
\colhead{$\dot{M}$$^{\rm f}$}&
\colhead{$\frac{1}{2}\dot{M}v^2$$^{\rm f}$}\\
\colhead{}&
\colhead{[$M_{\odot}~{\rm yr}^{-1}$]}&
\colhead{[$M_{\odot}~{\rm yr}^{-1}$]}&
\colhead{[\ergs]}&
\colhead{[$M_{\odot}~{\rm yr}^{-1}$]}&
\colhead{[\ergs]}}
\startdata
C~{\sc iv}    		& $0.9$   & \nodata      	& \nodata 				& \nodata 	& \nodata\\
Fe~{\sc xxvi} 		& \nodata & $220f_{\rm cov}$  	& $1.1 \times 10^{46}f_{\rm cov}$ 	& \nodata 	& \nodata\\
Fe~{\sc xvii} 		& \nodata & $8900f_{\rm cov}$ 	& $7.9 \times 10^{47}f_{\rm cov}$ 	& \nodata	& \nodata\\
S~{\sc xiv} 		& \nodata & $1700f_{\rm cov}$   & $1.6 \times 10^{47}f_{\rm cov}$ 	& \nodata	& \nodata\\
(c)   			& \nodata & \nodata       	& \nodata &$0.18f_{\rm cov}$ 		& $6.7 \times 10^{40}f_{\rm cov}$ \\
N~{\sc vi}$^{\rm d}$    & \nodata & \nodata       	& \nodata &$0.32f_{\rm cov}$ 		& $4.0 \times 10^{41}f_{\rm cov}$ \\
C~{\sc vi}$^{\rm d}$    & \nodata & \nodata       	& \nodata &$0.48f_{\rm cov}$ 		& $1.4 \times 10^{42}f_{\rm cov}$ \\
\enddata
\tablecomments{
(a) Ion from which the rates are computed.
(b) For a covering factor of 10\%.
(c) Best-fit outflow velocity found in \S \ref{xstar_calc1} of
1100 \kms.
(d) Using the velocities shown by the lines in Table \ref{tbl7}
(and $N_{\rm H}=10^{21.35}$ \cmn), of $\approx $ 2000 \kms, and 3000 \kms \sp
for N,
and C line respectively.
(e) From \cite{gibson2005a}.
(f) Present work.
}
\end{deluxetable}
\clearpage



\begin{figure}
\rotatebox{0}{\resizebox{15cm}{!}
{\plotone{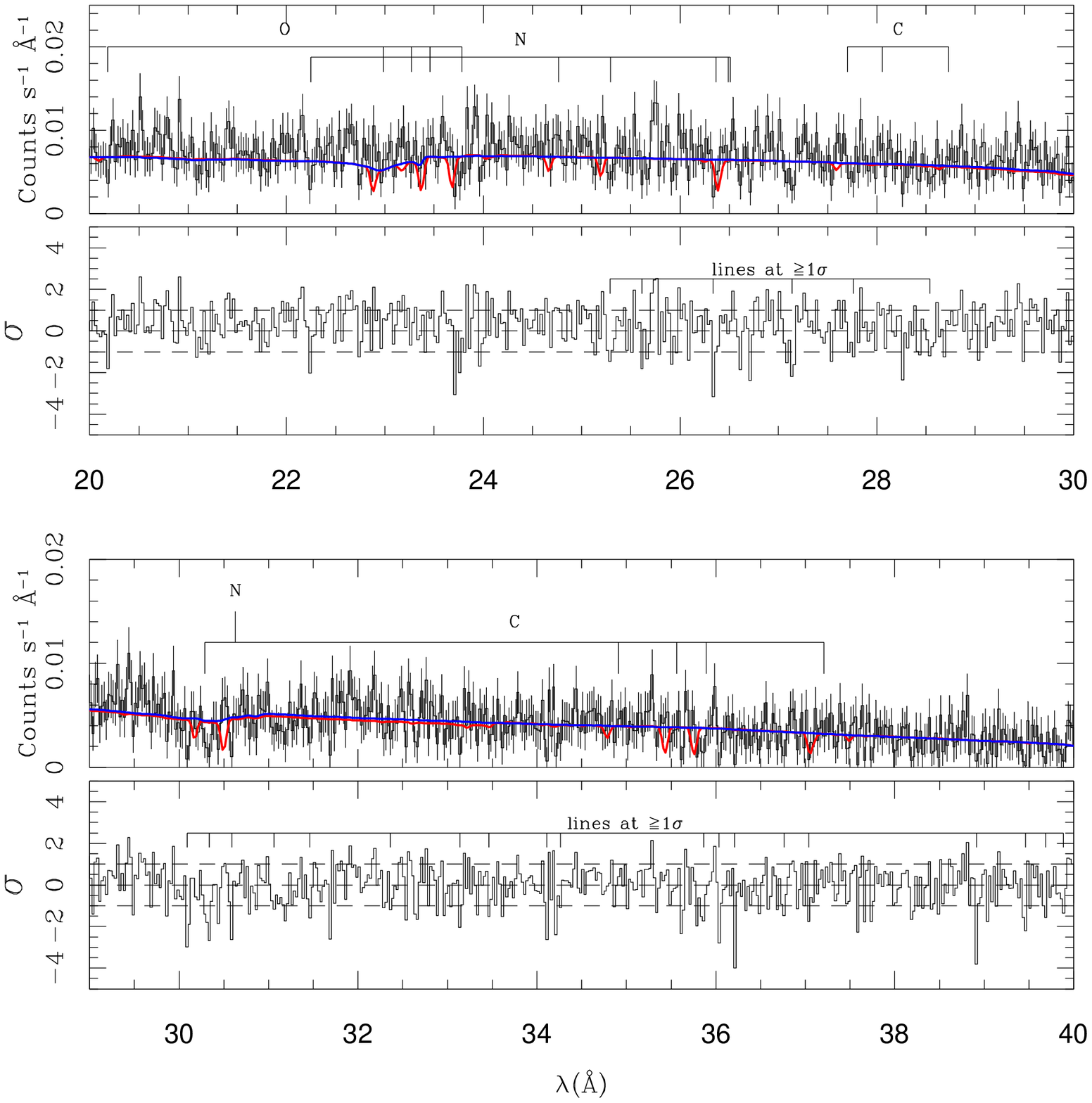}}}
\caption{
Comparison of the spectrum of \mr \sp with our
best-fit XSTAR photoionization model which is presented in high resolution
(bin size 25 m\AA). The physical model is shifted by $-1100$ \kms. Predicted lines
of Ne, Fe, O, C and N are marked as labels in the top of the spectrum.
\label{utas4}}
\end{figure}

\end{document}